\documentclass[aps]{revtex4}
\usepackage{amssymb,epsf}

\begin{document}

\title{Charged Dilatonic Black Holes in Gravity's Rainbow}
\author{S. H. Hendi$^{1,2}$\footnote{%
email address: hendi@shirazu.ac.ir}, Mir Faizal$^{3}$\footnote{%
email address: f2mir@uwaterloo.ca}, B. Eslam Panah$^{1}$ \footnote{%
email address: behzad.eslampanah@gmail.com} and S. Panahiyan$^{1,4}$ \footnote{%
email address: sh.panahiyan@gmail.com}} \affiliation{$^1$Physics
Department and Biruni Observatory, College of Sciences, Shiraz
University, Shiraz 71454, Iran\\
$^2$ Research Institute for Astronomy and Astrophysics of Maragha (RIAAM),
Maragha, Iran \\
$^3$ Department of Physics and Astronomy, University of Waterloo, Waterloo,
Ontario, N2L 3G1, Canada\\
$^4$Physics Department, Shahid Beheshti University, Tehran 19839, Iran}

\begin{abstract}
In this paper, we present charged dilatonic black holes in
gravity's rainbow. We study geometric and thermodynamic properties
of black hole solutions. We also investigate the effects of
rainbow functions on different thermodynamic quantities for these
charged black holes in dilatonic gravity's rainbow. Then, we
demonstrate that first law of thermodynamics is valid for these
solutions. After that, we investigate thermal stability of the
solutions using canonical ensemble and analyze the effects of
different rainbow functions on thermal stability. In addition, we
present some arguments regarding the bound and phase transition
points in context of geometrical thermodynamics. We also study the
phase transition in extended phase space in which cosmological
constant is treated as the thermodynamic pressure. Finally, we use
another approach to calculate and demonstrate that obtained
critical points in extended phase space are representing a second
order phase transition for these black holes.
\end{abstract}

\maketitle

\section{Introduction}

Motivated by works on Lifshitz scaling in condensed matter
physics, it is possible to take different Lifshitz scaling for
space and time, and the resultant theory is called Horava-Lifshitz
gravity \cite{HoravaPRD,HoravaPRL}. In the IR limit this gravity
reduces to general relativity, and so it can be considered as a UV
completion of general relativity. Motivated by this work on
Horava-Lifshitz gravity, different Lifshitz scaling for space and
time have been considered for type IIA string theory
\cite{Gregory2010}, type IIB string theory \cite{Burda2014},
AdS/CFT correspondence \cite
{Gubser2009,Ong2011,Alishahiha2014,Kachru2014}, dilatonic black
branes \cite {Goldstein2010,Bertoldi2010}, and dilatonic black
holes \cite{Kord Zangeneh2015, Tarrio2011}. Another UV completion
theory of general relativity which reduces to general relativity
in the IR limit is called gravity's rainbow \cite{MagueijoCQG}. In
fact, it has been demonstrated that gravity's rainbow is related
to Horava-Lifshitz gravity \cite{Garattini2015}. This is because
both of these theories are based on the modifying the usual
energy-momentum dispersion relation in the UV limit such that it
reduces to the usual energy-momentum dispersion relation in the IR
limit. It may be noted that such modification of the usual
energy-momentum has also been obtained in discrete spacetime
\cite{Hooft}, spacetime foam \cite {Camelia}, spin-network in loop
quantum gravity (LQG) \cite{Gambini}, ghost condensation
\cite{FaizalJPA} and non-commutative geometry \cite
{Carroll,FaizalMPLA}. The non-commutative geometry occurs due to
background fluxes in string theory \cite{st,st1}, and is used to
derive one of the most important rainbow functions in gravity's
rainbow \cite{Amelino,Jacob}.

It may be noted that the UV modification of the usual
energy-momentum relation implies the breaking of the Lorentz
symmetry in the UV limit of the theory. The spontaneous breaking
of the Lorentz symmetry can occur in string theory because of the
existence of an unstable perturbative string vacuum \cite{58}. It
is possible for a tachyon field to have wrong sign for its mass
squared in string field theory, and this causes the perturbative
string vacuum to become unstable. The theory becomes ill-defined
if the vacuum expectation value of the tachyon field is infinite.
It is also possible for the vacuum expectation value of the
tachyon field to be finite and negative. In this case, the
coefficient of the quadratic term for the massless vector field is
nonzero and negative, and this breaks the Lorentz symmetry. The
spontaneous breaking of the Lorentz symmetry in string theory has
also been investigated using the gravitational version of the
Higgs mechanism \cite{59}. This has been done for the low energy
effective action obtained from string theory. The Lorentz symmetry
breaking has been also studied using black brane in type IIB
string theory \cite{60}. In this analysis, moduli stabilization
was studied using a KKLT-type moduli potential in the context of
type IIB warped flux compactification. It was demonstrated that a
Higgs phase for gravity will exist if all moduli are stabilized.
Another study regarding the breaking of Lorentz symmetry was done
by using compactification in string field theory \cite{61}. It may
be noted that various other approaches to quantum gravity also
indicate that the Lorentz symmetry might only be an effective
symmetry which occurs in the IR limit of some fundamental theories
of quantum gravity \cite{Iengo,Arkani,Gripaios,Alfaro,Belich}.

Hence, there is a good motivation to study the UV deformation of geometries
that occur in string theory. In fact, motivated by Lifshitz deformation of
such geometries, and the relation between Horava-Lifshitz gravity and
gravity's rainbow \cite{Garattini2015}, recently rainbow deformation of
geometries that occur in string theory has been performed. Thus, the
modifications of the thermodynamics of black rings has been analyzed using
gravity's rainbow \cite{AliJHEP}. It has been observed that a remnant exists
for black rings in gravity's rainbow. It has also been argued that a remnant
might exist for all black objects in gravity's rainbow \cite{AliNPB}. This
has been explicitly demonstrated for Kerr black holes, Kerr-Newman black
holes in de Sitter space, charged AdS black holes, higher dimensional
Kerr-AdS black holes and black saturn \cite{AliNPB}. This was done by
generalizing the work done on the thermodynamics of black holes in gravity's
rainbow \cite{AliPRD2014}. As the usual uncertainty principle still holds in
gravity's rainbow \cite{LingMPLA,LingCQG}, and it is possible to obtain a
lower bound on the energy $E\geq 1/\Delta x$, using the usual uncertainty
principle. This energy can be related to the energy of a particle emitted in
Hawking radiation. Furthermore, the value of the uncertainty in position can
be equated to the radius of the event horizon, $E\geq 1/{\Delta x}\approx 1/{%
r_{+}}.$ This energy can be related to the energy at which
spacetime is probed, and hence it describes the energy $E$ in
gravity's rainbow. This is because effectively this particle
emitted with the energy $E$ can be viewed as a probe which is
probing the geometry of the black hole. This consideration
modifies the temperature of the black hole \cite{AliPRD2014}. The
entropy and heat capacity of black hole in gravity's rainbow can
be calculated using this modified temperature. An interesting
consequence of this modified solution is that it predicts the
existence of remnants for the black hole. Thus, the temperature of
the black hole reduces to zero, when black hole has a small but
finite size. At this size, the black hole does not emit any
Hawking radiation. The existence of black hole remnants can be
used as a solution for the information paradox
\cite{AliEPL,Gim2015}. Furthermore, it also solves problem related
to the existence of naked singularity at the last stage of the
evaporation of a black hole. As in this picture, a black hole does
not evaporate completely producing a naked singularity, but rather
a remnant is produced at the last stage of the evaporation of the
black hole. The existence of a remnant also has phenomenological
consequences. This is because it is not possible to produce black
holes smaller than these remnants. This increases the energy at
which mini black holes can be produced at the LHC \cite{AliPLB}.
Recently, a lot of interest has been generated in gravity's
rainbow \cite
{Garattini2014,HendiF2015,Chang2015,GarattiniII2014,Santos2015,Ali2015}.
It may be noted that the rainbow functions have been constrained
from experimental data \cite{Ali2015}. Black hole solutions in
gravity's rainbow with nonlinear sources have been investigated in
\cite {HendiPEMnonlinear}. In addition, the hydrostatic
equilibrium equation for this gravity was obtained in Ref.
\cite{HendiBEPTOV}.

As there is a strong motivation to study rainbow deformation of
geometries that occur in string theory, so we analyze the rainbow
deformation of charged dilatonic black holes in this paper. It may
be noted that dilaton gravity arises as a low-energy effective
field theory of string theory \cite {String1,String2}. The dilaton
field is also a candidate for dark matter \cite{DarkM}. In fact,
in order to have better picture of nature of the dark energy, a
new scalar field is added to the field content of the original
theory \cite{DarkE1,DarkE2}. Black objects in presence of dilaton
gravity, have also been investigated
\cite{dilaton1,dilaton2,dilaton3}. Recently, dilaton field has
been used for analyzing compact objects and hydrostatic
equilibrium of starts \cite{dilatonTOV1,dilatonTOV2}. The
evaporation of quantum black holes has also been investigated
using two dimensional dilaton gravity \cite{Evap1,Evap2}.
Motivated by these applications, we analyze dilaton field using
the formalism of gravity's rainbow.

Thermodynamical aspects of black holes have been of a great
interest ever since of pioneering works of Hawking and Beckenstein
\cite {Hawking,Beckenstein}. The idea that geometrical aspects of
black holes could be interpreted as thermodynamical quantities
provides a deep insight into the connection between gravity and
quantum mechanics. On the other hand, introduction and
developments in gauge/gravity duality highlighted the importance
of black holes thermodynamics \cite
{1AdSCFT1,AdSCFT1,AdSCFT2,AdSCFT3,AdSCFT4,AdSCFT5,AdSCFT6,AdSCFT7,AdSCFT8,AdSCFT9,AdSCFT10,AdSCFT11,2AdSCFT1,3AdSCFT1,4AdSCFT1}.
In addition, Hawking and Page showed the existence of a phase
transition for asymptotically anti de-Sitter black holes
\cite{Page}. This phase transition was reconsidered through the
use of AdS/CFT correspondence by Witten \cite {Witten}. These
works motivated a large number of researches to be conducted in
context of black holes thermodynamics, stability and their phase
transitions \cite{Thermodynamics1,Thermodynamics2,Thermodynamics3,
Thermodynamics4,Thermodynamics5,Thermodynamics6,Thermodynamics7,Thermodynamics8,Thermodynamics9,Thermodynamics10}.

Recently, it has been demonstrated that, it is possible to treat
the cosmological constant as the thermodynamic pressure in
extended phase space. There are several reasons for such
consideration which among them one can point out the existence of
second order phase transition for black holes, Van der Waals like
liquid/gas behavior in phase diagrams and formation of the triple
point \cite
{CosmP1,CosmP2,CosmP3,CosmP4,CosmP5,CosmP6,CosmP7,CosmP8,CosmP9,CosmP10,CosmP11,CosmP12,CosmP13,CosmP14,CosmP15,CosmP16,CosmP17,CosmP18,CosmP19}.
The consideration of the cosmological constant as a
thermodynamical variable could be supported by studies that are
conducted in context of AdS/CFT
\cite{CosVar2,CosVar3,CosVar4,CosVar5}. In addition, it was shown
that a case of ensemble dependency exists for charged
$3$-dimensional black holes which could be removed by considering
cosmological constant as a thermodynamical variable
\cite{Mamasani}. The thermodynamical critical behavior of black
holes in presence of different matter fields and gravities has
been investigated in literature \cite
{CosmP1,CosmP2,CosmP3,CosmP4,CosmP5,CosmP6,CosmP7,CosmP8,CosmP9,CosmP10,CosmP11,CosmP12,CosmP13,CosmP14,CosmP15,CosmP16,CosmP17,CosmP18,CosmP19}

Another interesting method of studying thermodynamical structure
of the black holes is through the use of geometry. It is proposed
that one can build a phase space of the black holes by employing
one of thermodynamical quantities of the black holes as
thermodynamical potential and its corresponding extensive
parameters as components of the phase space. The information
regarding phase transitions of the black holes is within the
singularities of Ricci scalar of the constructed phase space. In
other words, the divergencies of the Ricci scalar of
thermodynamical metric are representing bound and phase transition
points. The thermodynamical potential for this method could be
mass which is used in Weinhold \cite {Weinhold1,Weinhold2},
Quevedo \cite{Quevedo1,Quevedo2,Quevedo3} and HPEM
\cite{HPEM1,HPEM2,HPEM3} metrics or entropy which is employed in
Ruppeiner \cite{Ruppeiner1,Ruppeiner2} metric. It was pointed out
that Ruppeiner and Weinhold methods are related to each other with
temperature as conformal factor \cite{Quevedo1,Quevedo2,Quevedo3}.
It was shown that for specific cases of black holes, Weinhold,
Ruppeiner and Quevedo metrics may fail to provide consistent
results regarding phase transitions while the HPEM metric is
proven to be successful one \cite{HPEM1,HPEM2,HPEM3}. In what
follows, we use all of these methods to study phase transitions of
the black holes.

This paper is organized as follows. We obtain the charged black
hole solutions in dilaton gravity's rainbow and analyze their
properties. This will be done by making the metric of charged
black hole solutions in dilaton gravity depends on the energy. We
also examine the first law of thermodynamics for this solution.
Next, we study the stability of such solutions in gravity's
rainbow and phase transition of these black holes through heat
capacity, geometrical thermodynamics and the analogy between
cosmological constant and thermodynamical pressure. Finally, we
obtain critical pressure and horizon radius through another
method. Last section is devoted to conclusion.

\section{Charged dilatonic black hole solutions in Gravity's Rainbow}

In this section, we obtain charged black hole solutions in dilaton gravity's
rainbow and investigate their properties. This will be done by writing an
energy dependent version of the metric for dilaton-Maxwell gravity. It may
be noted that gravity's rainbow is based on the generalization of doubly
special relativity \cite{MagueijoPRD}, and so it is not possible for a
particle to attain energy greater than the Planck energy in gravity's
rainbow. This is because gravity affects particles of different energies
differently, and so the spacetime is represented by a family of energy
dependent metrics in gravity's rainbow \cite{MagueijoCQG}. The gravity's
rainbow can be constructed by considering the following deformation of the
standard energy-momentum relation
\begin{equation}
E^2f^{2}(\varepsilon)-p^2g^{2}(\varepsilon)=m^2,  \label{MDR}
\end{equation}
where energy ratio is $\varepsilon=E/E_{P}$, in which $E$ and $E_P$ are,
respectively, the energy of test particle and the Planck energy. The
functions $f(\varepsilon)$ and $g(\varepsilon)$ are required to be
constrained in such a way that the standard energy-momentum relation is
obtained in the infrared limit. Thus, we require
\begin{equation}
\lim\limits_{\varepsilon\to0} f(\varepsilon)=1,\qquad
\lim\limits_{\varepsilon \to0} g(\varepsilon)=1.
\end{equation}

It may be noted that the spacetime is probed at the energy $E$,
and by definition this cannot be greater than the Planck energy. $
f^{2}(\varepsilon) $ and $g^{2}(\varepsilon)$ are called the
rainbow functions and their functional forms are
phenomenologically motivated. Now it is possible to define an
energy dependent deformation of the metric $\hat{g}(\varepsilon)$
as \cite{Peng}
\begin{equation}
\hat{g}(\varepsilon)=\eta^{ab}e_a(\varepsilon)\otimes e_b(\varepsilon),
\label{rainmetric}
\end{equation}
where
\begin{equation}
e_0(\varepsilon)=\frac{1}{f(\varepsilon)}\tilde{e}_0, \qquad
e_i(\varepsilon)=\frac{1}{g(\varepsilon)}\tilde{e}_i,
\end{equation}
here $\tilde{e}_0$ and $\tilde{ e}_i$ refer to the energy independent frame
fields.

The $4$-dimensional action of charged dilaton gravity is \cite{Chan}
\begin{equation}
\mathcal{I}=\frac{1}{16\pi }\int d^{4}x\sqrt{-g}\left[ \mathcal{R}-2\left(
\nabla \Phi \right) ^{2}-V\left( \Phi \right) -e^{-2\alpha \Phi }F_{\mu \nu
}F^{\mu \nu }\right] ,  \label{action}
\end{equation}%
where $\mathcal{R}$ is the Ricci scalar curvature, $\Phi $ is the dilaton
field and $V\left( \Phi \right) $ is a potential for $\Phi $. The
electromagnetic field is $F_{\mu \nu }=\partial _{\mu }A_{\nu
}-\partial_{\nu }A_{\mu }$ in which $A_{\mu }$ is the electromagnetic
potential. In addition, it should be pointed out that $\alpha $ is a
constant which determines the strength of coupling of the scalar and
electromagnetic field. Due to the fact that we are looking for the black
hole with a radial electric field ($F_{tr}(r)=-F_{rt}(r) \neq 0$), the
electromagnetic potential will be in the following form
\begin{equation}
A_{\mu }=\delta _{\mu }^{0}h\left( r\right) .  \label{electric po}
\end{equation}

Using variational principle and varying Eq. (\ref{action}) with respect to
the gravitational field $g_{\mu \nu }$, the dilaton field $\Phi $ and the
gauge field $A_{\mu }$, we can obtain the following field equations
\begin{equation}
R_{\mu \nu }=2\left( \partial _{\mu }\Phi \partial _{\nu }\Phi +\frac{1}{4}%
g_{\mu \nu }V(\Phi )\right) +2e^{-2\alpha \Phi }\left( F_{\mu \eta }F_{\nu
}^{\eta }-\frac{1}{4}g_{\mu \nu }F_{\lambda \eta }F^{\lambda \eta }\right) ,
\label{dilaton equation(I)}
\end{equation}%
\begin{equation}
\nabla ^{2}\Phi =\frac{1}{4}\frac{\partial V}{\partial \Phi }-\frac{\alpha }{%
2}e^{-2\alpha \Phi }F_{\lambda \eta }F^{\lambda \eta },
\label{dilaton equation(II)}
\end{equation}%
\begin{equation}
\nabla _{\mu }\left( e^{-2\alpha \Phi }F^{\mu \nu }\right) =0.
\label{Maxwell equation}
\end{equation}

In this paper we are attempting to obtain dilaton-Maxwell rainbow solutions.
To do so, one can employ following static metric ansatz
\begin{equation}
ds^{2}=-\frac{\Psi (r)}{f^{2}(\varepsilon)}dt^{2}+\frac{1}{g^{2}(\varepsilon)%
}\left[ \frac{dr^{2}}{\Psi (r)}+r^{2}R^{2}(r)d\Omega _{k}^{2}\right] ,
\label{metric}
\end{equation}%
where $\Psi (r)$ and $R(r)$ are radial dependent functions which
should be determined, and $d\Omega _{k}^{2}$ represents the line
element of a $2-$ dimensional hypersurface with the constant
curvature $2k$ and volume $\varpi _{2}$. We should note that the
constant $k$ indicates that the boundary of $t=constant$ and
$r=constant$ can be a positive (elliptic), zero (flat) or negative
(hyperbolic), constant curvature hypersurface with following
explicit forms
\begin{equation}
d\Omega _{k}^{2}=\left\{
\begin{array}{cc}
d\theta ^{2}+\sin ^{2}\theta d\varphi ^{2}, & k=1 \\
d\theta ^{2}+\sinh ^{2}\theta d\varphi ^{2}, & k=-1 \\
d\theta ^{2}+d\varphi ^{2}, & k=0%
\end{array}%
\right. .  \label{dOmega}
\end{equation}

Using Eq. (\ref{Maxwell equation}), one can obtain electromagnetic tensor as
\begin{equation}
F_{tr}=\frac{qe^{2\alpha \Phi }}{ r^{2}R(r)^{2} },  \label{Ftr eq}
\end{equation}%
where $q$ is an integration constant which is related to the electric charge
of the black hole.

Here, in order to find consistent metric functions, we use a modified
version of Liouville-type dilation potential with following form
\begin{equation}
V(\Phi )=\frac{2k\alpha ^{2}}{b^{2}\mathcal{K}_{-1,1}}g^{2}(\varepsilon)e^{%
\frac{2\Phi }{\alpha }}+2\Lambda e^{2\alpha \Phi },  \label{V(Phi)}
\end{equation}%
where $\mathcal{K}_{i,j}=i+j\alpha ^{2}$ and $\Lambda $ is a free
parameter which plays the role of the cosmological constant. It is
worthwhile to mention that for the case of
$g(\varepsilon)=f(\varepsilon)=1$, one obtains
\begin{equation}
\lim_{g(\varepsilon)=f(\varepsilon)\rightarrow 1}V(\Phi )=\frac{2k\alpha ^{2}%
}{b^{2}\mathcal{K}_{-1,1}}e^{\frac{2\Phi }{\alpha }}+2\Lambda e^{2\alpha
\Phi },
\end{equation}%
which is the usual Liouville-type dilation potential that is used
in the context of Friedman-Robertson-Walker scalar field
cosmologies \cite {Ozer} and Einstein-Maxwell-dilaton black holes
\cite{Chan,Yazadjiev,Sheykhi}.

Next, we employ an ansatz, $R(r)=e^{\alpha \Phi (r)}$, in the
field equations. The motivation for considering such an ansatz is
due to black string solutions of Einstein-Maxwell-dilaton gravity
which was first introduced in Ref. \cite{Dehghani}. Now, we are in
a position to obtain metric functions. It is a matter of
calculation to show that by using Eq. (\ref{Ftr eq}), the metric
(\ref{metric}) and the mentioned ansatz for $R(r)$ , we have
following solutions for the field equations, (Eqs. (\ref{dilaton
equation(I)}) and (\ref{dilaton equation(II)}))
\begin{equation}
\Psi(r)=-\frac{\mathcal{K}_{1,1}}{\mathcal{K}_{-1,1}}\left( \frac{b}{r}%
\right) ^{-2\gamma }k-\frac{m}{r^{\frac{\mathcal{K}_{1,-1}}{\mathcal{K}_{1,1}%
}}}+\frac{\mathcal{K}_{1,1}^{2}\Lambda r^{2}}{g^{2}(\varepsilon )\mathcal{K}%
_{-3,1}}\left( \frac{b}{r}\right) ^{2\gamma }
+\frac{q^{2}\mathcal{K}_{1,1}f^{2}(\varepsilon )}{r^{2}}\left(
\frac{b}{r} \right) ^{-2\gamma },  \label{f(r)}
\end{equation}%
\begin{equation}
\Phi (r)=\frac{\alpha }{\mathcal{K}_{1,1}}\ln \left( \frac{b}{r}\right) ,
\label{Phi(r)}
\end{equation}%
where $b$ is an arbitrary constant and $\gamma =\alpha ^{2}/\mathcal{K}%
_{1,1} $. In the above expression, $m$ is an integration constant which is
related to the total mass of the black hole. It is notable that, in the
absence of a non-trivial dilaton ($\alpha =\gamma =0$), the solution (\ref%
{f(r)}) reduces to
\begin{equation}
\Psi (r)=k-\frac{m}{r}-\frac{\Lambda }{3}\frac{r^{2}}{g^{2}(\varepsilon )}+%
\frac{f^{2}(\varepsilon )q^{2}}{r^{2}},
\end{equation}%
which describes a $4$-dimensional asymptotically AdS topological charged
black hole in gravity's rainbow with a positive, zero or negative constant
curvature hypersurface.

In order to confirm black hole interpretation of the solutions, we
look for the curvature singularity. To do so, we calculate the
Kretschmann scalar. Calculations show that for finite values of
radial coordinate, the Kretschmann scalar is finite. On the other
hand, for very small and very large values of $r$, we obtain
\begin{eqnarray}
\lim_{r\rightarrow 0}R_{\alpha \beta \mu \nu }R^{\alpha \beta \mu \nu }
&\propto &r^{-\frac{4\mathcal{K}_{2,1}}{\mathcal{K}_{1,1}}},  \label{RR0} \\
\lim_{r\rightarrow \infty }R_{\alpha \beta \mu \nu }R^{\alpha \beta \mu \nu
} &=&\frac{12\Lambda (\alpha ^{4}-2\alpha ^{2}+2)}{\mathcal{K}_{3,-1}^{2}}%
\left( \frac{b}{r}\right) ^{4\gamma }.  \label{RRinf}
\end{eqnarray}

\begin{figure}[tbp]
$%
\begin{array}{ccc}
\epsfxsize=5.5cm \epsffile{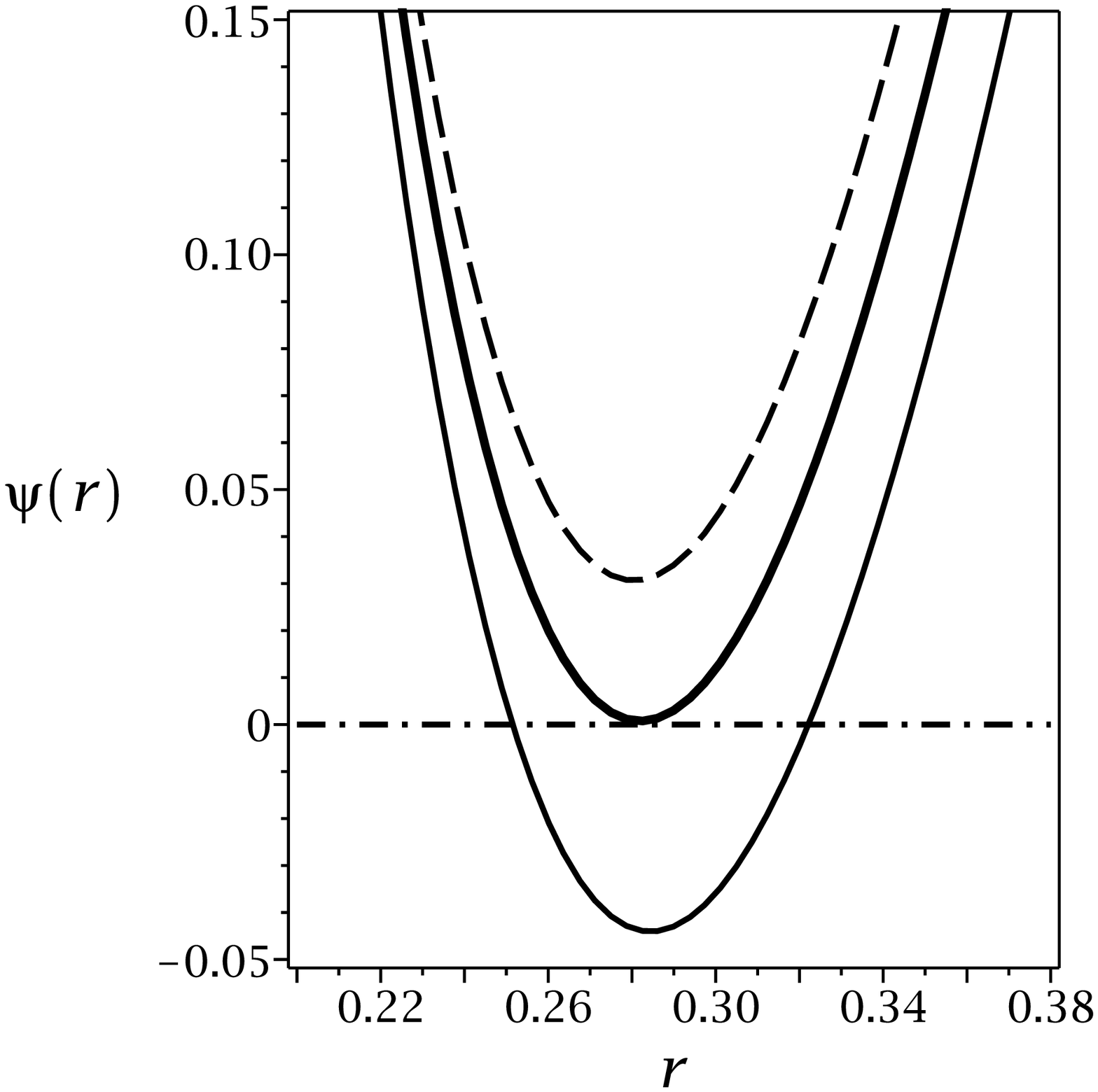} & \epsfxsize=5.5cm %
\epsffile{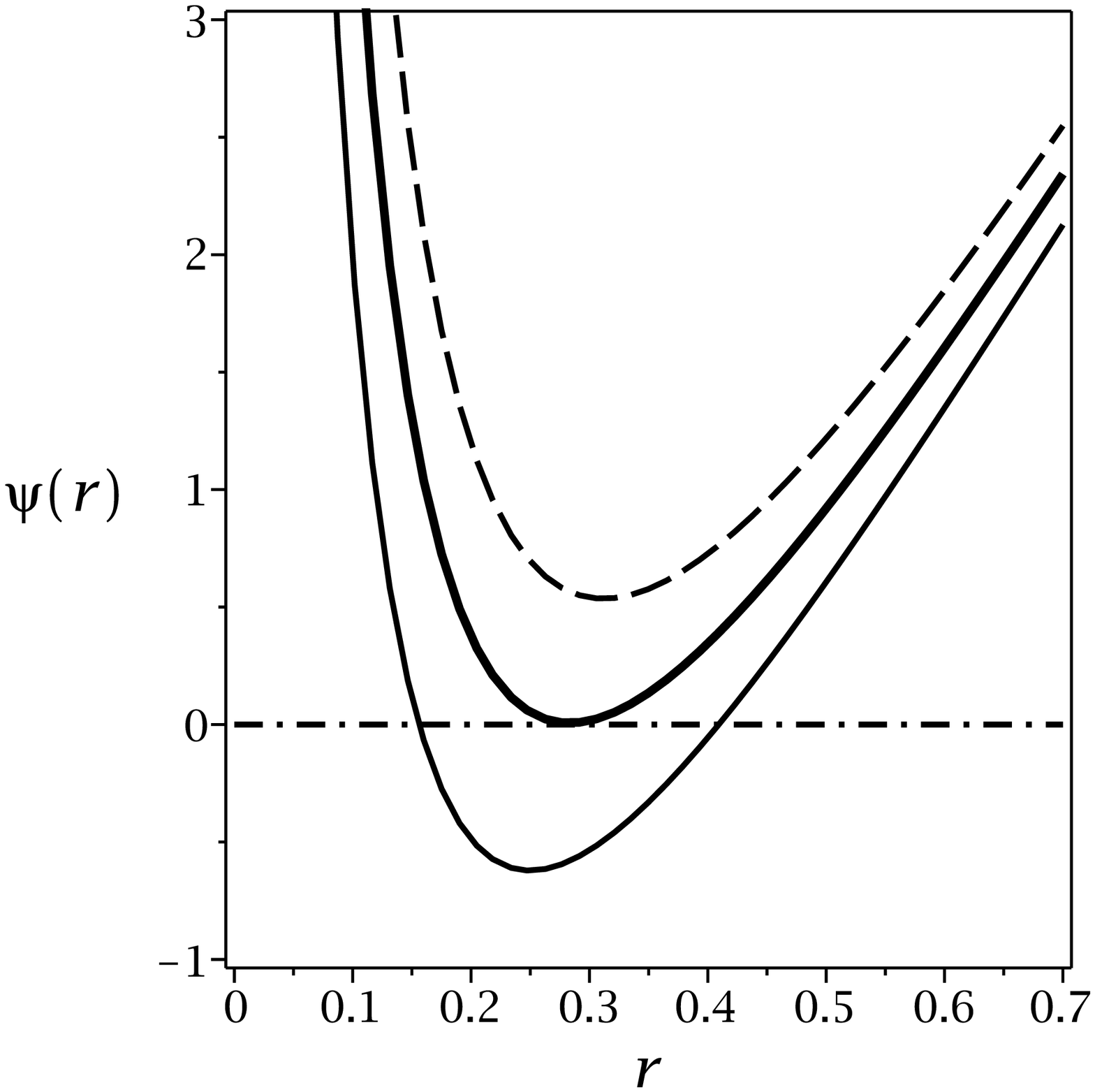} & \epsfxsize=5.5cm \epsffile{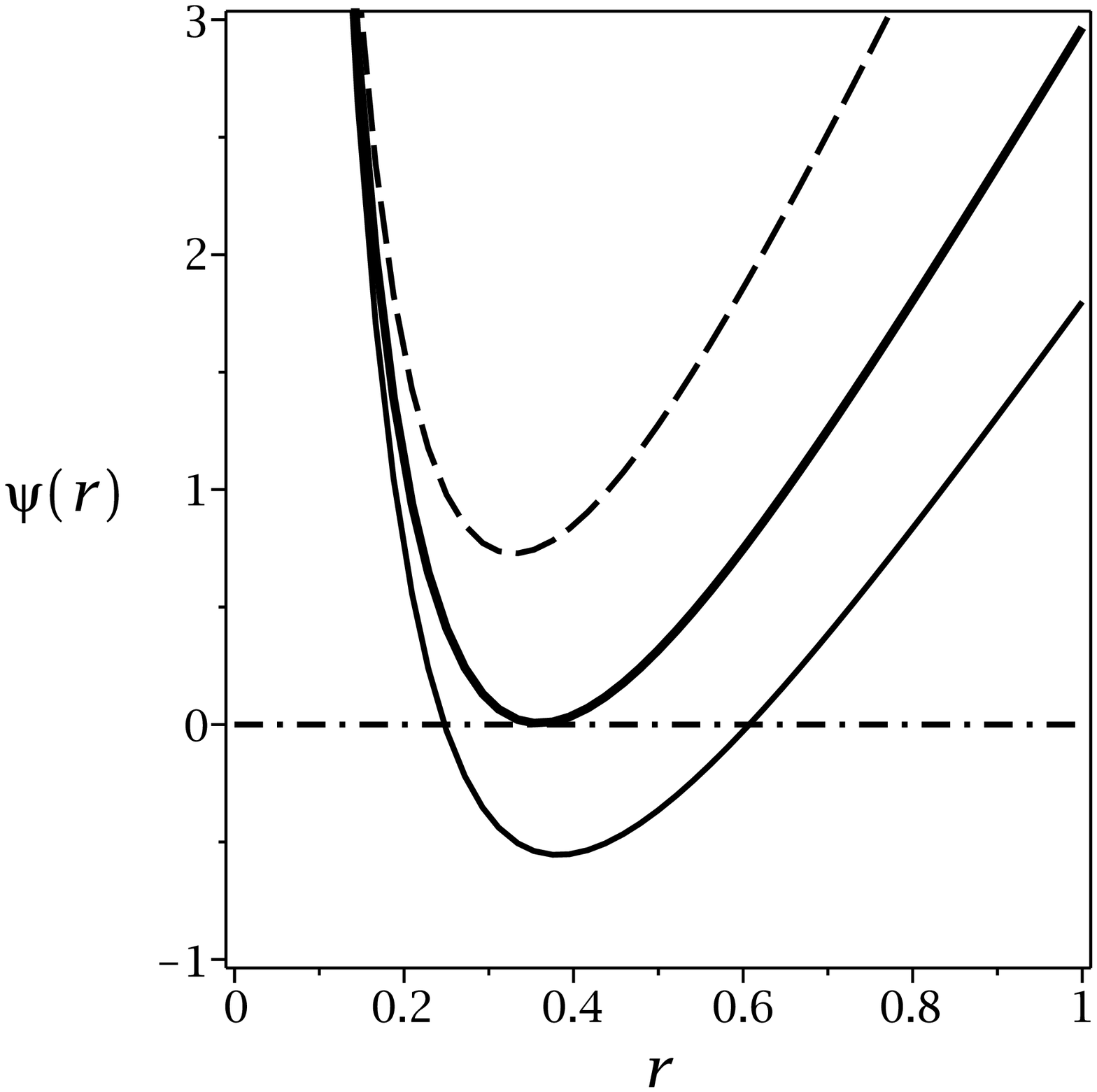}%
\end{array}
$%
\caption{$\Psi(r)$ versus $r$ for $k=1$, $m=5$, $\Lambda=-0.5$, $b=1.2$ and $%
q=0.68$.\newline
Left panel for $\protect\alpha=0.9$, $f^{2}(\protect\varepsilon)=1$, $g^{2}(%
\protect\varepsilon)=0.85$ (dashed line), $g^{2}(\protect\varepsilon)=0.96$
(bold line) and $g^{2}(\protect\varepsilon)=1.20$ (continuous line).\newline
Middle panel for $\protect\alpha=0.9$, $g^{2}(\protect\varepsilon)=1$, $%
f^{2}(\protect\varepsilon)=0.80$ (dashed line), $f^{2}(\protect\varepsilon%
)=1.05$ (bold line) and $f^{2}(\protect\varepsilon)=1.20$ (continuous line).%
\newline
Right panel for $g^{2}(\protect\varepsilon)=1.3$, $f^{2}(\protect\varepsilon%
)=1.3$, $\protect\alpha=0.85$ (dashed line), $\protect\alpha=0.87$ (bold
line) and $\protect\alpha=0.9$, (continuous line).}
\label{metricFig}
\end{figure}

Equation (\ref{RR0}) confirms that there is an essential singularity located
at $r=0$, while Eq. (\ref{RRinf}) shows that for nonzero $\alpha $, the
asymptotical behavior of the solutions is not AdS. It is easy to show that
the metric function may contain real positive roots (see Fig. \ref{metricFig}%
), and therefore, the curvature singularity can be covered with an event
horizon and interpreted as a black hole.

\section{Thermodynamical Quantities}

Now, we are in a position to calculate thermodynamic and conserved
quantities of obtained solutions and examine the validity of the first law
of thermodynamics.

In order to obtain the temperature, we use the concept of surface
gravity to show that the temperature of these solutions has
following form
\begin{equation}
T=-\frac{g(\varepsilon )\mathcal{K}_{1,1}\left( \frac{b}{r_{+}}\right)
^{-2\gamma }}{4\pi }\left[ \frac{q^{2}f(\varepsilon )}{r_{+}^{3}}+\frac{%
r_{+}\Lambda }{g^{2}(\varepsilon )f(\varepsilon )}\left( \frac{b}{r_{+}}%
\right) ^{4\gamma }+\frac{k}{f(\varepsilon )r_{+}\mathcal{K}_{-1,1}}\right] .
\label{temp}
\end{equation}

On the other hand, one can use the area law for extracting modified version
of the entropy related to the Einsteinian class of black objects with the
following structure
\begin{equation}
S=\frac{\varpi _{2}r_{+}^{2}}{4g^{2}(\varepsilon)}\left( \frac{b}{r_{+}}%
\right) ^{2\gamma },  \label{entropy}
\end{equation}%
in which by setting $\alpha=0$ and $%
g(\varepsilon)=1 $, the entropy of Einstein-Maxwell-dilation black
holes is recovered. In order to find the total electric charge of
the solutions, one can use the Gauss law. Calculating the flux of
electric field helps us to find the total electric charge with the
following form
\begin{equation}
Q=\frac{\varpi _{2}qf(\varepsilon)}{4\pi g(\varepsilon)}.  \label{Q}
\end{equation}

Next, we are interested in obtaining the electric potential. Using following
standard relation, one can obtain the electric potential at the event
horizon with respect to the infinity as a reference
\begin{equation}
U\left( r\right) =\left. A_{\mu }\chi ^{\mu }\right\vert _{r\longrightarrow
\infty }-\left. A_{\mu }\chi ^{\mu }\right\vert _{r\longrightarrow r_{+}}=-%
\frac{2q}{r_{+}^{3}}.
\end{equation}

Finally, according to the definition of mass due to Abbott and Deser \cite%
{AD1,AD2,AD3}, the total mass of the solution is
\begin{equation}
M=\frac{\varpi _{2}b^{2\gamma }}{8\pi \mathcal{K}_{1,1}g(\varepsilon)f(%
\varepsilon)}m.  \label{Mass}
\end{equation}

It is worthwhile to mention that for limiting case of $g(\varepsilon)=f(%
\varepsilon)=1$ and $\alpha=0$, Eq. (\ref{Mass}) reduces to mass
of the Einstein-Maxwell black holes \cite{Sheykhi}. In addition,
in obtained conserved and thermodynamical quantities, only the
electric potential remains unaffected by considering gravity's
rainbow.

Now, we are in a position to check the validity of the first law
of thermodynamics. To do so, first, we calculate the geometrical
mass, $m$, by using $f\left( r=r_{+}\right) =0$. Then by employing
obtained relation for geometrical mass and Eq. (\ref{Mass}) for
total mass of the black holes, we find
\begin{equation}
M(r_{+},q)=\frac{\varpi _{2}Ab^{2\gamma }}{8\pi \mathcal{K}%
_{1,-1}g(\varepsilon )f(\varepsilon )},  \label{Mass2}
\end{equation}%
where%
\[
A=\left( \frac{b}{r_{+}}\right) ^{2\gamma }\left[ kr_{+}^{\frac{\mathcal{K}%
_{1,-1}}{\mathcal{K}_{1,1}}}-q^{2}f^{2}(\varepsilon )\mathcal{K}%
_{-1,1}r_{+}^{\frac{-\mathcal{K}_{1,3}}{\mathcal{K}_{1,1}}}+\frac{\Lambda
\mathcal{K}_{1,1}\mathcal{K}_{1,-1}}{g^{2}(\varepsilon )\mathcal{K}_{-3,1}}%
r_{+}^{\frac{\mathcal{K}_{3,1}}{\mathcal{K}_{1,1}}}\left( \frac{b}{r_{+}}%
\right) ^{4\gamma }\right] .
\]

It is a matter of calculation to show that
\begin{equation}
\left( \frac{\partial M}{\partial S}\right) _{Q}=T\text{ \ \ \ \ }\&\text{\
\ \ \ \ }\left( \frac{\partial M}{\partial Q}\right) _{S}=U.
\end{equation}

Therefore, we proved that the first law is valid as
\begin{equation}
dM=\left( \frac{\partial M}{\partial S}\right) _{Q}dS+\left( \frac{\partial M%
}{\partial Q}\right) _{S}dQ.
\end{equation}

\section{Thermal stability}

In this section, we study thermal stability of the solutions in
context of the canonical ensemble. The stability conditions in the
context of canonical ensemble are determined by the sign of the
heat capacity. In other words, the positivity/negativity of the
heat capacity is denoted as the black object being in
stable/unstable state. Therefore, in order to study the stability
of the charged black holes in dilatonic gravity's rainbow, we
study the changes in the sign of corresponding heat capacity. It
is worthwhile to mention that investigating the behavior of the
heat capacity enables one to obtain the phase transitions of the
solutions at the same time. The root and divergence point of the
heat capacity are denoted as bounded point ($r_{+0}$) and second
order phase transition point ($r_{+c}$), respectively. Bounded
point is related to the root of temperature and the sign of $T$
changes at $r_{+0}$, while we expect to obtain positive
temperature at $r_{+c}$.

The system in canonical ensemble is considered to be in fixed
charge. Therefore, we have
\begin{equation}
C_{Q}=\left( \frac{\partial M}{\partial S}\right) _{Q}\left( \frac{\partial
^{2}M}{\partial S^{2}}\right) _{Q}^{-1}.  \label{CQ}
\end{equation}

Considering the mentioned bounded and phase transition points, one can
obtain
\begin{equation}
\left\{
\begin{array}{cc}
T=\left( \frac{\partial M}{\partial S}\right) _{Q}=0 & bounded\text{ } point
\\
&  \\
\left( \frac{\partial T}{\partial S}\right) _{Q}=\left( \frac{\partial ^{2}M%
}{\partial S^{2}}\right) _{Q}=0 & phase\text{ } transition \text{ }point%
\end{array}%
\right. .  \label{phase}
\end{equation}

There are three valuable known cases for the rainbow functions
which are characteristics of the rainbow solutions. These three
cases are arisen from different phenomenological origins with an
upper limit for considering the energy of test particle $E$
\[
\varepsilon=\frac{E}{E_{P}}\leq 1.
\]

The first case is originated from loop quantum gravity and non-commutative
geometry. In this case, we have following relations for rainbow functions of
metric \cite{Amelino,Jacob}
\begin{equation}
f(\varepsilon)=1,\text{ \ \ }g(\varepsilon) =\sqrt{1-\eta \varepsilon ^{n}}.
\label{loop}
\end{equation}

The other case is constructed by considering the hard spectra from gamma-ray
bursts which leads to \cite{Camelia}
\begin{equation}
f(\varepsilon)=\frac{e^{\beta \varepsilon}-1}{\beta \varepsilon},\text{ \ \ }%
g(\varepsilon) =1.  \label{gamma}
\end{equation}

Interestingly, opposite to previous case, in this one the effect of $%
g(\varepsilon) $ which is coupled to spatial coordinates of the metric is
vanished, whereas in the case one which is related to loop quantum gravity
the effect of coupling term for time component of the metric is vanished.

Finally, in the case three, the choices of rainbow functions are due to
constancy of the velocity of the light \cite{Magueijo}
\begin{equation}
f( \varepsilon) =g(\varepsilon) =\frac{1}{1-\lambda \varepsilon}.
\label{light}
\end{equation}

Using first law of thermodynamics, one can rewrite the relation for heat
capacity into
\begin{equation}
C_{Q}=T\left( \frac{\partial S}{\partial r_{+}}\right) _{Q}\left( \frac{%
\partial T}{\partial r_{+}}\right) _{Q}^{-1}.  \label{CQQ}
\end{equation}

Now, by employing Eqs. (\ref{temp}) and (\ref{entropy}) with (\ref{CQQ}),
one can show that heat capacity is%
\begin{equation}
C_{Q}=\frac{r_{+}^{2}\left( \frac{b}{r_{+}}\right) ^{2\gamma }\left( q^{2}%
\mathcal{K}_{-1,1}g^{2}(\varepsilon )f^{2}(\varepsilon
)+r_{+}^{2}kg^{2}(\varepsilon )+r_{+}^{4}\Lambda \mathcal{K}_{-1,1}\left(
\frac{b}{r_{+}}\right) ^{4\gamma }\right) }{2g^{2}(\varepsilon )\mathcal{K}%
_{-1,1}\left( r_{+}^{2}kg^{2}(\varepsilon )-q^{2}\mathcal{K}%
_{3,1}g^{2}(\varepsilon )f^{2}(\varepsilon )-r_{+}^{4}\Lambda \mathcal{K}%
_{-1,1}\left( \frac{b}{r_{+}}\right) ^{4\gamma }\right) }.  \label{heat}
\end{equation}

Considering obtained relation for the heat capacity (Eq.(\ref{heat})) and
three mentioned cases for the rainbow functions of the metric (Eqs. (\ref%
{loop}-\ref{light})), we study the stability of the solutions. In case of
horizon flat ($k=0$), one can find that the root(s) of the heat capacity and
divergence point(s) are given by following relations
\begin{equation}
r_{+0}=b\left( -\frac{b^{4}\Lambda }{g^{2}(\varepsilon )f^{2}(\varepsilon
)q^{2}}\right) ^{-\frac{\mathcal{K}_{1,1}}{4}},  \label{root-flat}
\end{equation}%
\begin{equation}
r_{+c}=b\left( -\frac{b^{4}\Lambda \mathcal{K}_{-1,1}}{g^{2}(\varepsilon
)f^{2}(\varepsilon )q^{2}\mathcal{K}_{3,1}}\right) ^{-\frac{\mathcal{K}_{1,1}%
}{4}}.  \label{div}
\end{equation}

Interestingly, for horizon flat, the root of the heat capacity,
hence bound point is independent of the dilaton parameter, $\alpha
$ while divergency of the heat capacity is a function of this
parameter. On the other hand, in
order to have positive real valued divergency in AdS spacetime, we have $%
\alpha >1$ restriction for dilaton parameter. This condition for
dS spacetime is opposite. In other words, the real valued
divergency is obtained if $0\leq \alpha <1$. (see Eqs.
(\ref{root-flat}) and (\ref{div}) for more details). In order to
charged black holes in dilatonic gravity's rainbow with flat
horizon be stable, following conditions must be hold
\[
\left\{
\begin{array}{c}
-\Lambda r_{+}^{4}\left( \frac{b}{r_{+}}\right) ^{6\gamma }\leq
q^{2}g(\varepsilon )^{2}f(\varepsilon )^{2}\left( \frac{b}{r_{+}}\right)
^{2\gamma } \\
-\Lambda r_{+}^{4}\mathcal{K}_{-1,1}\mathcal{K}_{1,1}\left( \frac{b}{r_{+}}%
\right) ^{2\gamma }\leq q^{2}g(\varepsilon )^{2}f(\varepsilon )^{2}\mathcal{K%
}_{3,1}\left( \frac{b}{r_{+}}\right) ^{-2\gamma }%
\end{array}%
\right. .
\]

As for the cases of $k=\pm 1$, it was not possible to obtain the analytical
relations for the root and divergence point of the heat capacity. Therefore,
we employ numerical method for studying the properties of the heat capacity
for spherical and hyperbolic horizons. As for the stability conditions,
there are different orders of the horizon radius for each term. These terms
will have dominant effect in specific regions of horizon radius and other
parameters. Considering the effectiveness of these terms, the stability
conditions will vary from one case to another one. Also, these effective
behaviors may present different regions of stability and instability. Taking
a closer look at different terms, one can see that the most effective
parameter in stability conditions which modifies the exponent of horizon
radius highly and changes the positivity and negativity of each term, is
dilaton parameter, $\alpha$. In other words, considering different values of
$\alpha$, stable/unstable regions will be modified highly. This highlights
the effect of dilaton field on thermodynamical behavior of the solutions.

\begin{figure}[tbp]
$%
\begin{array}{cc}
\epsfxsize=5.5cm \epsffile{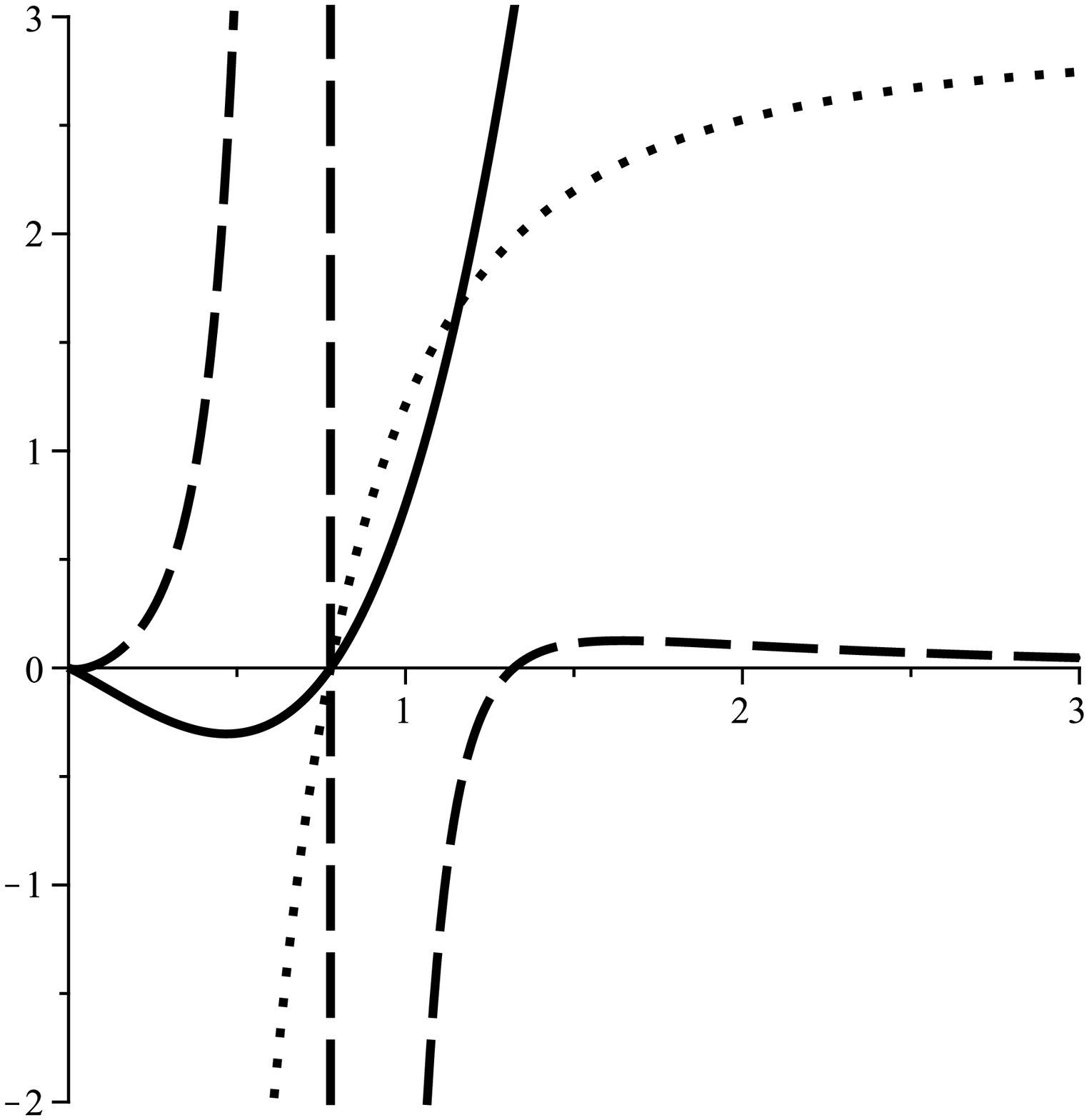} & \epsfxsize=5.5cm %
\epsffile{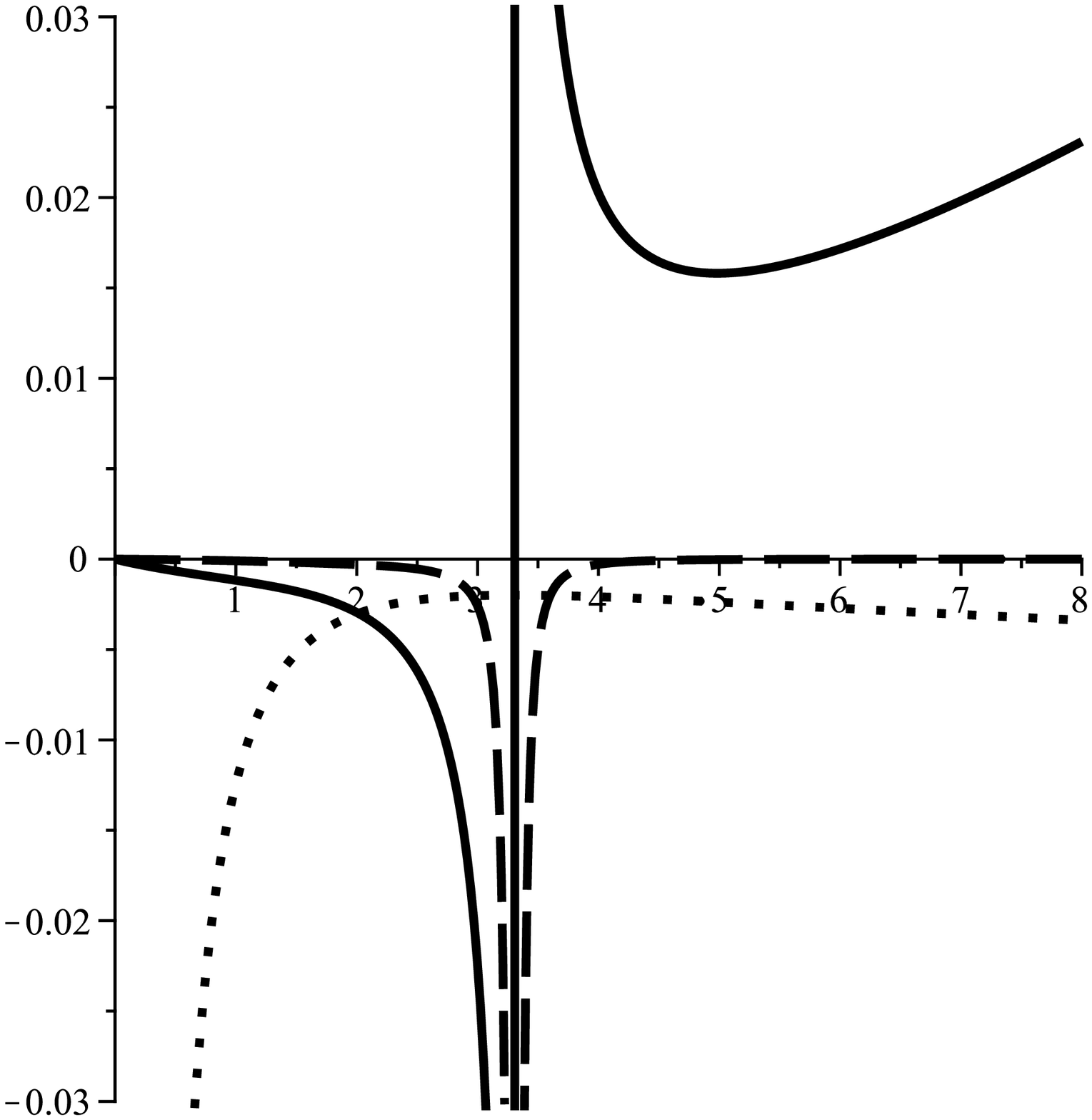}%
\end{array}
$%
\caption{$C_{Q}$ (continuous line), $T$ (dotted line) and
$\mathcal{R}$ (dashed line) versus $r_{+}$ for $k=1$,
$\Lambda=-1$, $b=5$, $E=1$, $E_{p}=5$ and $q=1$. \newline
$\protect\alpha=0.9$ (left panel) and $\protect\alpha=1.09$ (right
panel) "different scales".} \label{Fig1}
\end{figure}

\begin{figure}[tbp]
$%
\begin{array}{ccc}
\epsfxsize=5.5cm \epsffile{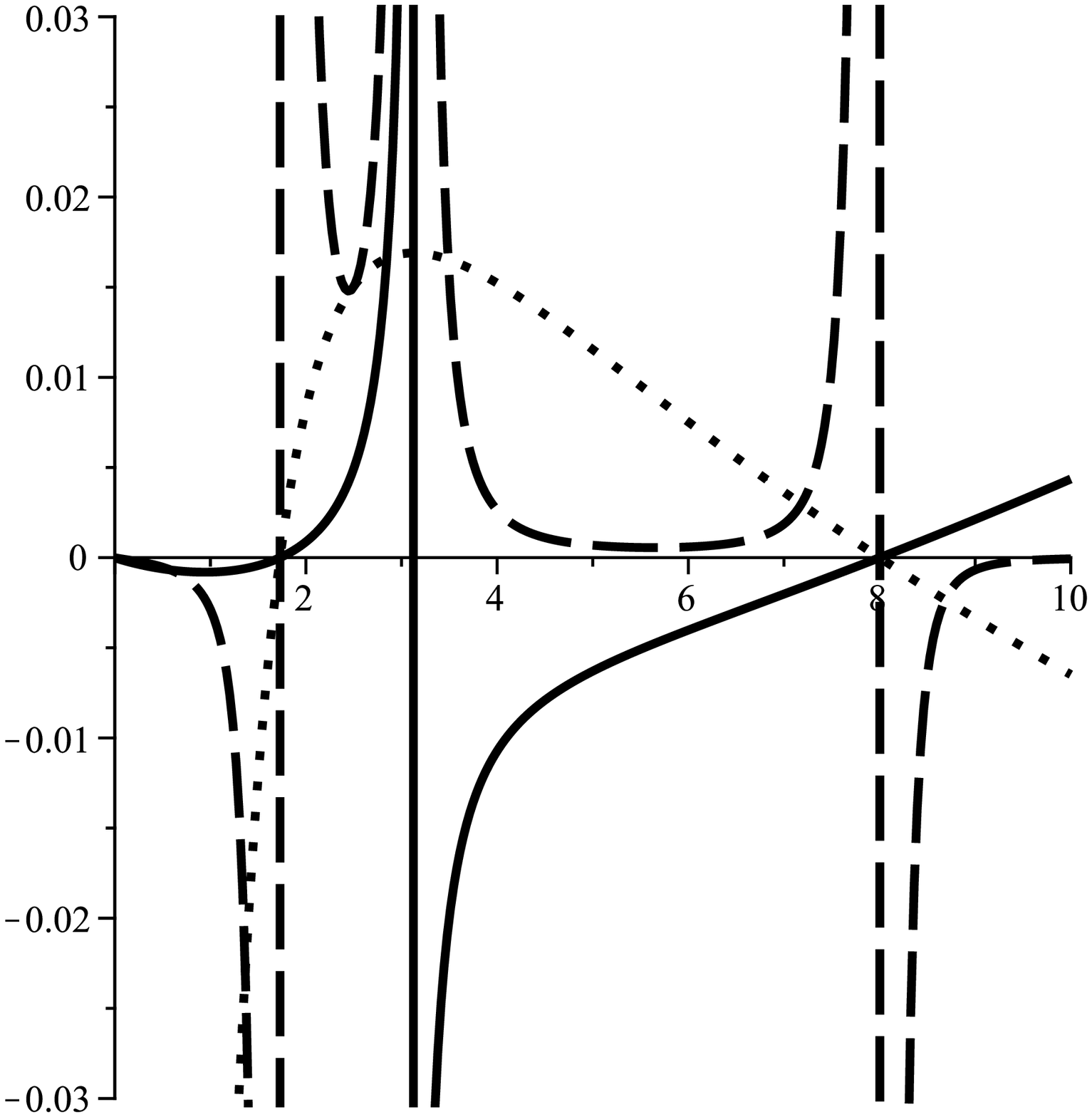} & \epsfxsize=5.5cm %
\epsffile{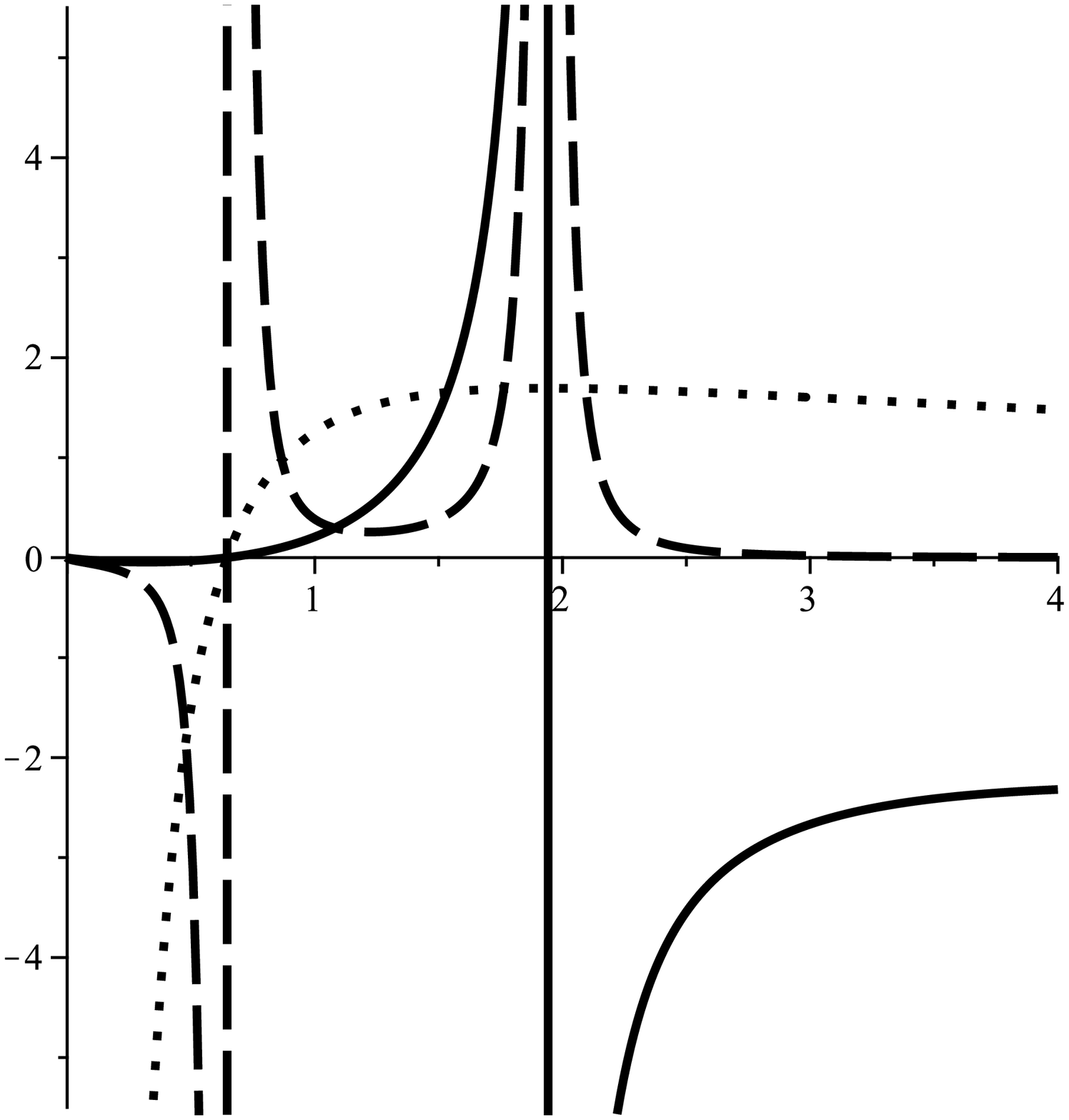} & \epsfxsize=5.5cm \epsffile{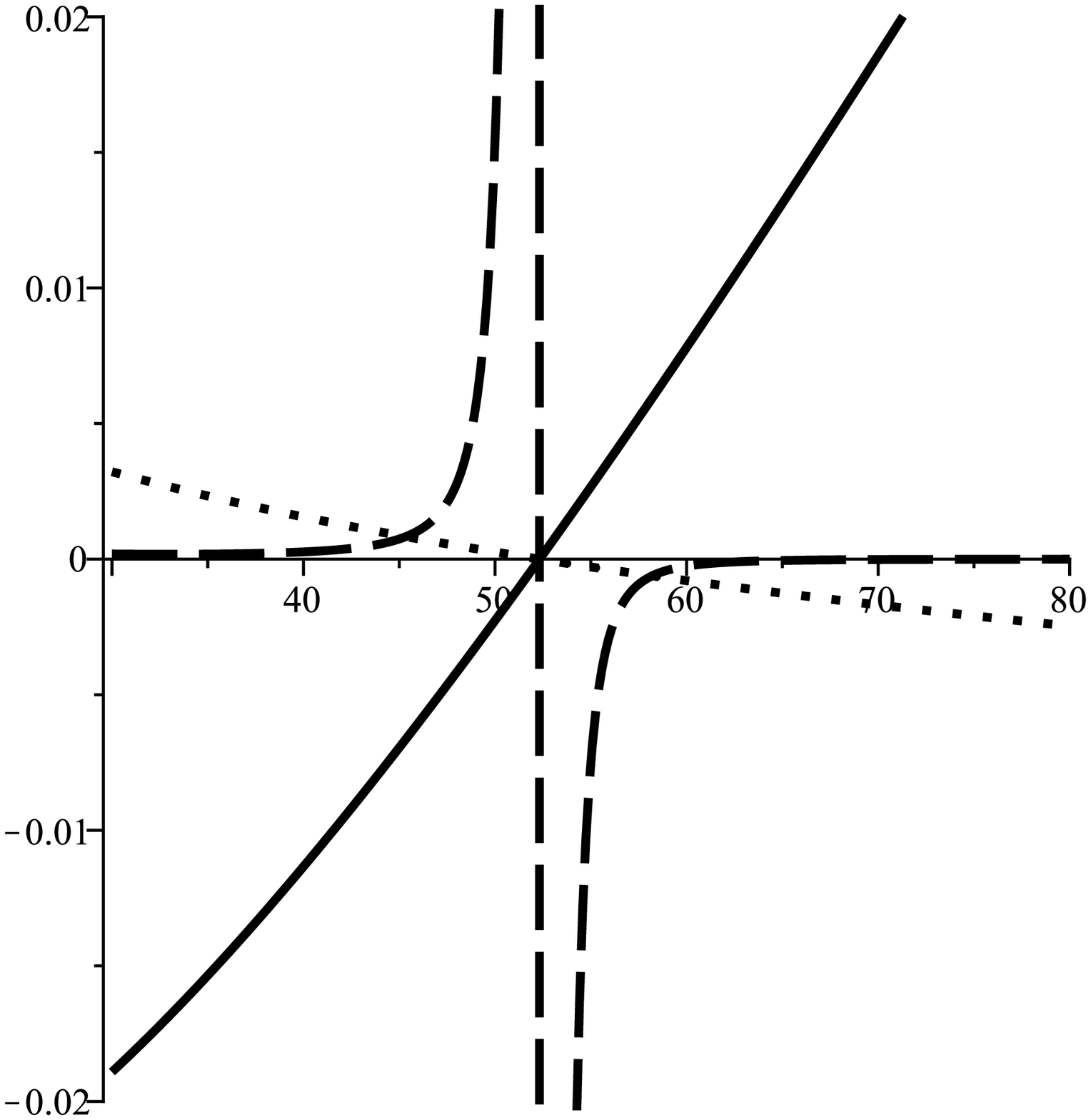}%
\end{array}
$%
\caption{$C_{Q}$ (continuous line), $T$ (dotted line) and
$\mathcal{R}$ (dashed line) versus $r_{+}$ for $k=1$,
$\Lambda=-1$, $b=5$, $E=1$, $E_{p}=5$ and $q=1$. \newline
$\protect\alpha=1.1$ (left panel) and $\protect\alpha=1.3$ (middle
and right panels) "different scales".} \label{Fig2}
\end{figure}

\begin{figure}[tbp]
$%
\begin{array}{cc}
\epsfxsize=5.5cm \epsffile{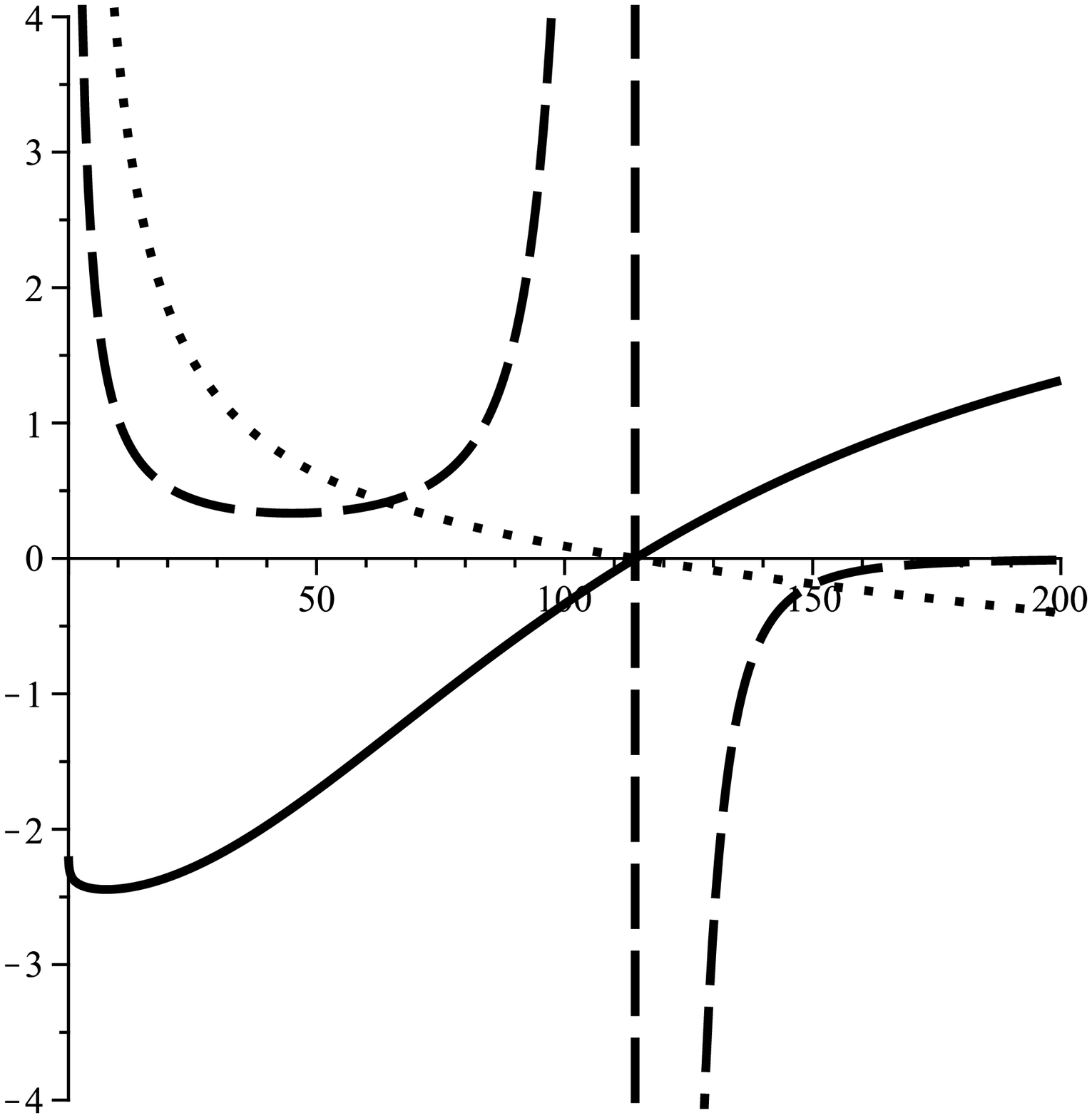} & \epsfxsize=5.5cm %
\epsffile{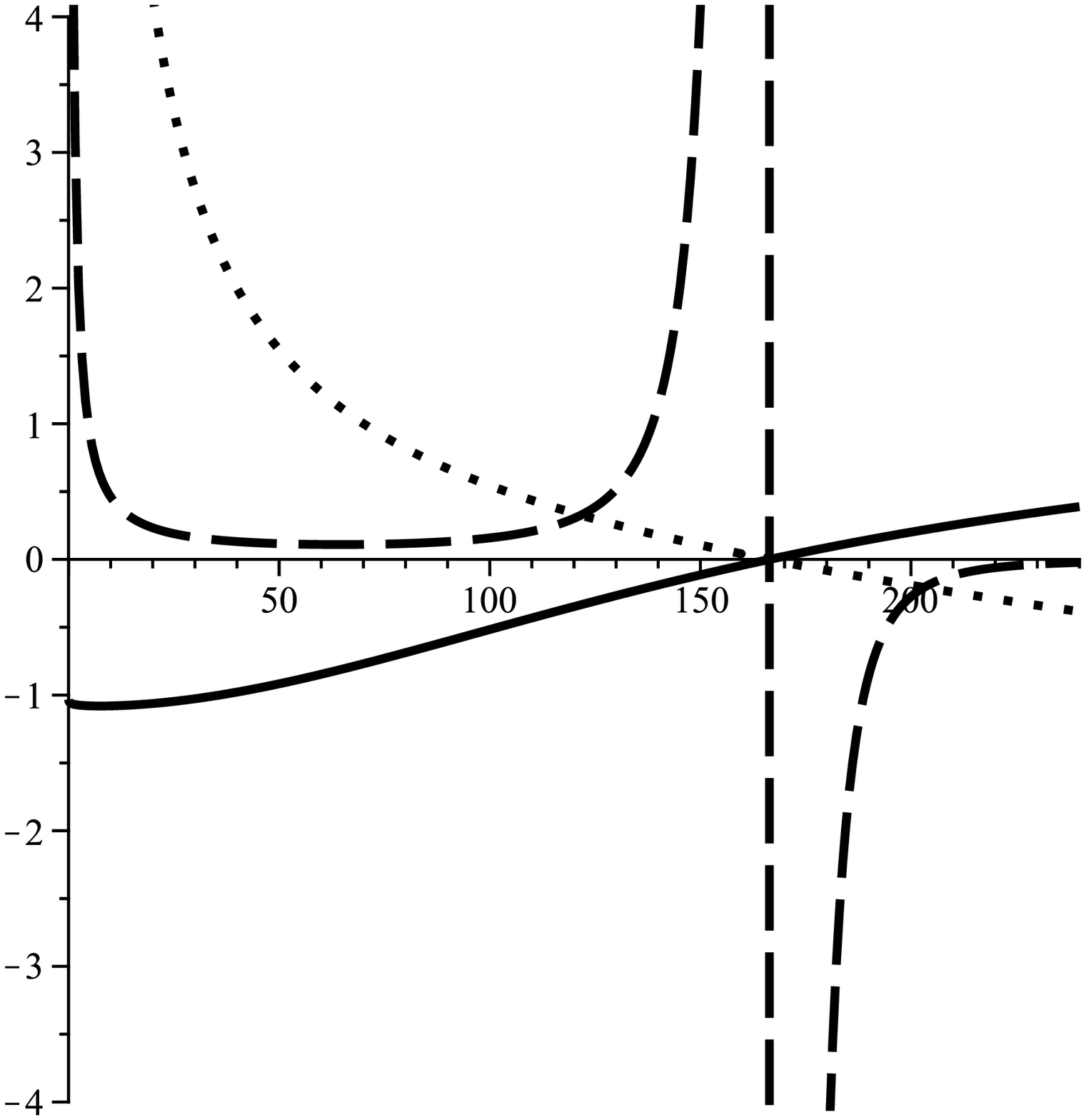}%
\end{array}
$%
\caption{$C_{Q}$ (continuous line), $T$ (dotted line) and
$\mathcal{R}$ (dashed line) versus $r_{+}$ for $k=1$,
$\Lambda=-1$, $b=5$, $E=1$, $E_{p}=5$ and $q=1$. \newline
$\protect\alpha=10$ (left panel) and $\protect\alpha=15$ (right
panel) "different scales".} \label{Fig3}
\end{figure}

In order to give a better insight regarding the thermodynamical
behavior of these black holes, we study the behavior of the
temperature. The reason is the fact that negativity of the
temperature represents nonphysical systems which are not of our
interest. Therefore, we study the conditions for
positivity/negativity of the temperature. Considering Eq.
(\ref{temp}), there are three terms which are related to electric
charge, cosmological constant and topological structure of the
metric. The effectiveness of each term is a function of their
factors. Therefore, considering different values for these factors
may lead to one of the following scenarios: one root and
temperature is an increasing function of horizon radius (left
panel of Fig. \ref{Fig1}), no root and temperature is negative
with one maximum (right panel of Fig. \ref{Fig1}), two roots with
one region of positivity and two regions of negativity (Fig.
\ref{Fig2}), one root and temperature is a decreasing function of
the horizon radius (Fig. \ref{Fig3}).

In general, charged term is always negative in Eq. (\ref{temp}).
If one considers AdS solutions, the second term will be positive.
As for the last term, if spherical solution is chosen, then in
order for topological term be positive, the dilaton parameter must
be $\alpha>1$. Whereas in case of hyperbolic horizon the condition
will be modified into $\alpha<1$. On the other hand, if dS
solutions are considered, the second term will be negative.
Therefore, the possibility of having positive temperature depends
on topological term with mentioned conditions for spherical and
hyperbolic cases. It is worthwhile to mention that horizon flat of
dS solutions has negative temperature. Therefore, it is not
physical. Considering the mentioned changes in different terms of
temperature, depending on the dominant regions of the different
each term, the temperature will have positive/negative value with
different behavior (which were pointed out in plotted diagrams).

\begin{figure}[tbp]
$%
\begin{array}{ccc}
\epsfxsize=5.5cm \epsffile{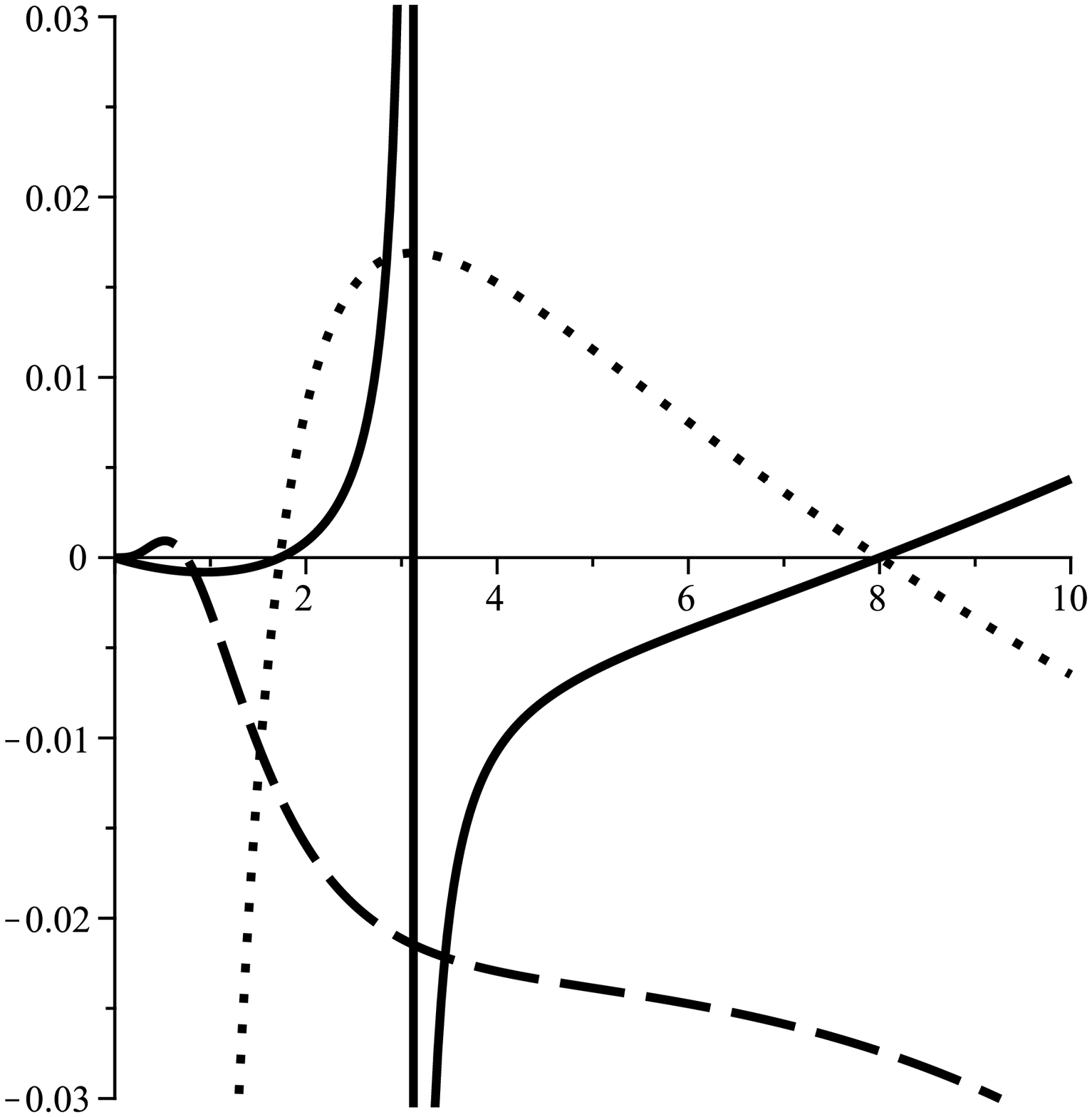} & \epsfxsize=5.5cm %
\epsffile{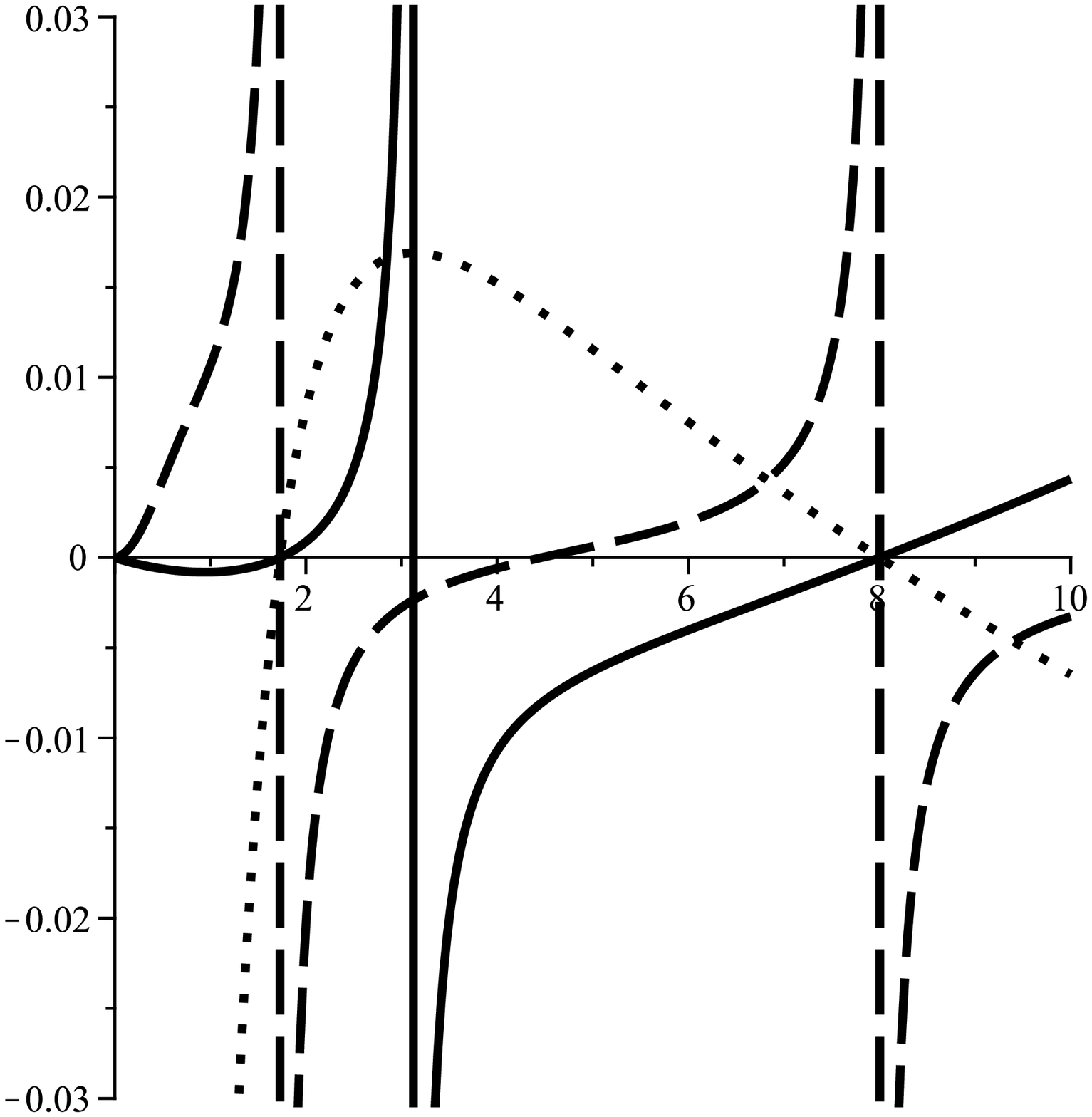} & \epsfxsize=5.5cm \epsffile{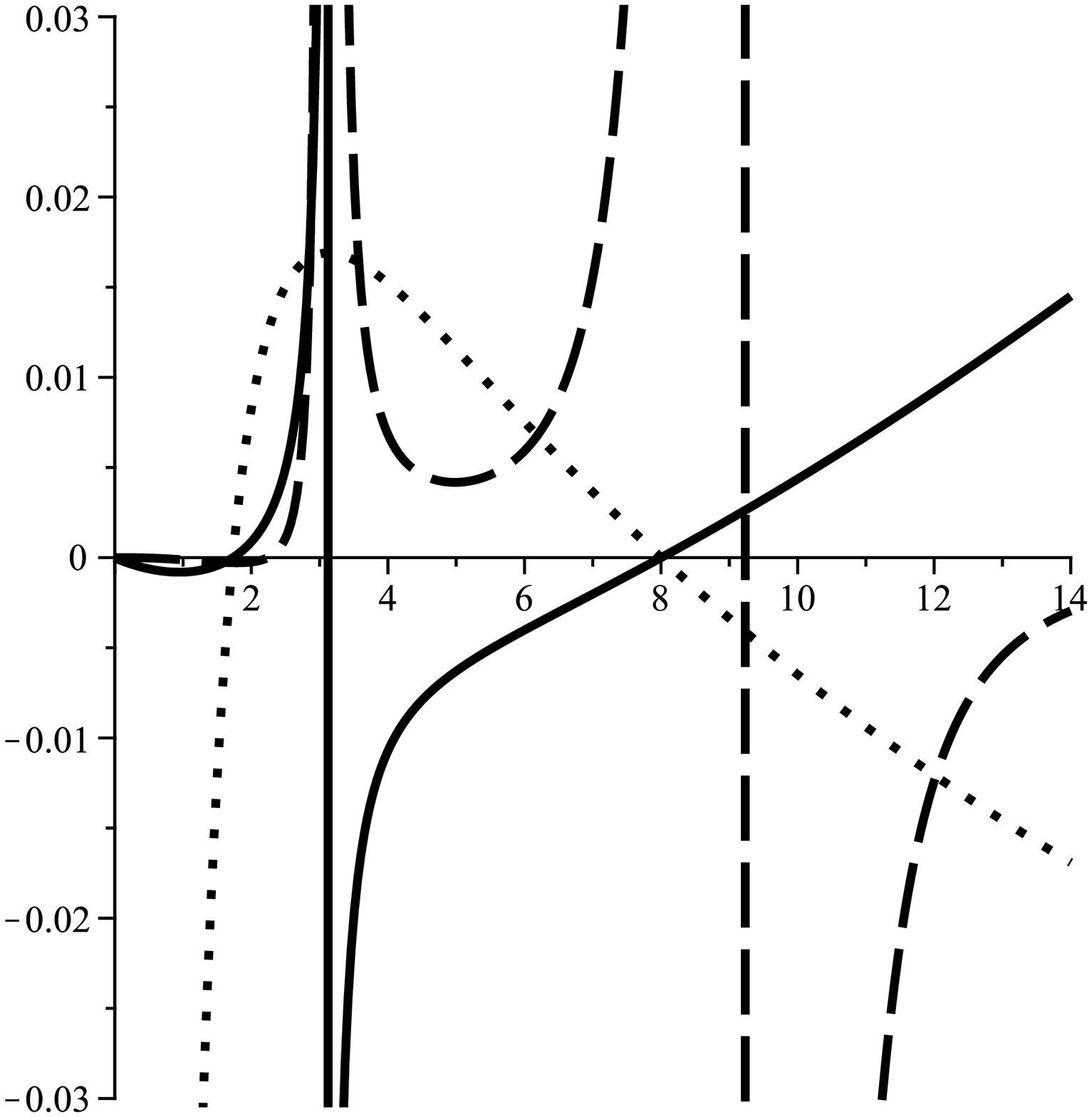}%
\end{array}
$%
\caption{$C_{Q}$ (continuous line), $T$ (dotted line) and
$\mathcal{R}$
(dashed line) versus $r_{+}$ for $k=1$, $\Lambda=-1$, $b=5$, $E=1$, $E_{p}=5$%
, $q=1$ and $\protect\alpha=1.1$. \newline Weinhold (left panel),
Ruppeiner (middle panel) and Quevedo (middle panel) "different
scales".} \label{Fig33}
\end{figure}


In order to elaborate the effects of the dilation field on thermal stability
and mentioned behaviors for the temperature, we plot various diagrams using
the first model of rainbow functions of gravity's rainbow.

It is evident that in case of temperature being an increasing
function of horizon radius, for positive temperature, we have
stable black holes and heat capacity is an increasing function of
$r_{+}$ (Fig. \ref{Fig1} left panel). Increasing the dilaton
parameter will lead to formation of two stable and unstable states
where both of these states are in negative temperature (Fig.
\ref{Fig1} right panel). On the other hand, by increasing dilaton
parameter, the temperature will have two regions of negativity and
one positivity. In the positive region, a phase transition takes
place between unstable larger state to smaller stable state. This
phase transition point is represented by a divergency of the heat
capacity. Increasing dilaton parameter leads to decreasing the
place of divergency of the heat capacity and increasing the region
in which we have unstable physical solutions (Fig. \ref{Fig2}).
The larger root of the temperature is highly sensitive to
variation of the $\alpha $ (see Fig. \ref{Fig2}). For sufficiently
large values of the dilaton parameter, the temperature is a
decreasing function of the horizon radius with one root. The heat
capacity in the region of the positive temperature is negative,
therefore in this case rainbow solutions are unstable (Fig.
\ref{Fig3}). It is worthwhile to mention that the root in this
case is an increasing function of dilaton parameter (compare two
diagrams of Fig. \ref{Fig3}).

\section{Geometrical thermodynamics}

In this section, we will conduct a study regarding the phase
transition points of the obtained black holes through the use of
geometrical thermodynamics concept. In this concept, one can
construct the thermodynamical structure of the black hole through
the use of thermodynamical variables. In other words, by using one
of the thermodynamical variable as potential and its corresponding
extensive parameters, it is possible to build phase space. The
singularities of the Ricci scalar of this phase space, marks two
different properties of the solutions: one is a bound point which
is related to the root of the temperature and marks the physical
and non-physical solutions. The other one is related to
singularities of the heat capacity which are marking the points
which system goes under phase transition.

Considering such property for the constructed phase space, a valid
approach of the geometrical thermodynamics produces a Ricci scalar
which has singularities that cover both of the mentioned points.
Depending on the thermodynamical potential, the extensive
parameters would be different for each phase space. One of these
potentials could be entropy which could be used to construct
Ruppeiner phase space. An other potential could be mass which is
employed to build Weinhold, Quevedo and HPEM phase spaces. The
mentioned phase spaces have following forms
\cite{HPEM1,HPEM2,HPEM3}
\begin{equation}
ds^{2}=\left\{
\begin{array}{cc}
Mg_{ab}^{W}dX^{a}dX^{b} & Weinhold \\ \\
-MTg_{ab}^{W}dX^{a}dX^{b} & Ruppeiner \\ \\\left(
SM_{S}+QM_{Q}\right) \left( -M_{SS}dS^{2}+M_{QQ}dQ^{2}\right) &
Quevedo \\ \\
S\frac{M_{S}}{M_{QQ}^{3}}\left( -M_{SS}dS^{2}+M_{QQ}dQ^{2}\right) & HPEM%
\end{array}%
\right. ,  \label{met}
\end{equation}%
where their corresponding denominator of their Ricci scalars are

\begin{equation}
denom(\mathcal{R})=\left\{
\begin{array}{cc}
\left( M_{SS}M_{QQ}-M_{SQ}^{2}\right) ^{2}M^{2}\left( S,Q\right) & Weinhold
\\ \\
\left( M_{SS}M_{QQ}-M_{SQ}^{2}\right) ^{2}T(S,Q)M^{2}\left(
S,Q\right) & Ruppeiner \\ \\
\left( SM_{S}+QM_{Q}\right) ^{3}M_{SS}^{2}M_{QQ}^{2} & Quevedo \\
\\
S^{3}M_{S}^{3}M_{SS}^{2} & HPEM%
\end{array}%
\right. ,  \label{Ric}
\end{equation}%
in which $M_{QQ}=\left( \frac{\partial ^{2}M}{\partial Q^{2}}\right) _{S}$, $%
M_{SQ}=\frac{\partial ^{2}M}{\partial S\partial Q}$, $M_{SS}=\left( \frac{%
\partial ^{2}M}{\partial S^{2}}\right) _{Q}$ and $M_{S}=\left( \frac{%
\partial M}{\partial S}\right) _{Q}$.

Now, by using Eqs. (\ref{entropy}), (\ref{Q}), (\ref{Mass2}) and (\ref{met}%
), one can construct mentioned phase spaces and calculate their
corresponding curvature scalar for these black holes. Due to
economical reasons, we will not present obtained relation for
Ricci scalar but demonstrate results in plotted diagrams (see
Figs. \ref{Fig1}-\ref{Fig3} for HPEM metric and Fig. \ref{Fig33}
for other metrics). Fig. \ref{Fig33} shows that for specific
values one can find cases in which Weinhold (left panel of Fig.
\ref{Fig33}), Ruppeiner (middle panel of Fig. \ref{Fig33}) and
Quevedo (right panel Fig. \ref{Fig33}) will not produce suitable
divergencies in their Ricci scalar to cover mentioned points. In
other words, the divergencies of their Ricci scalar may not
coincide with root and divergencies of the heat capacity. On the
other hand, it is seen that all the divergencies of the curvature
scalar of the HPEM metric match with bound and phase transition
points of the heat capacity. The nature of the behavior of the
Ricci scalar around each one of these divergencies enables one to
recognize whether it is a bound point or a divergence point in
heat capacity \cite{HPEM1,HPEM2,HPEM3}.

\section{Phase Transitions in Extended Phase space}

In this section, we investigate the existence of second order phase
transition through the analogy between negative cosmological constant and
thermodynamical pressure. The usual relation for pressure and cosmological
constant is give by \cite{CosmP9,CosmP10,CosmP11,Gunasekaran}
\begin{equation}
P=-\frac{\Lambda }{8\pi }.  \label{PPP}
\end{equation}

It was shown that the gravitational theory under consideration may
affect this relation and modifies it \cite{Armanfar,DehghaniKSH}.
In calculations of the conserved and thermodynamical quantities,
we found that these quantities were modified due to existence of
gravity's rainbow and dilaton field. It is natural to question
whether the usual relation between cosmological constant and
thermodynamical pressure could be modified in presence of the
dilaton field as well as rainbow functions. To investigate such
modification, we use the right hand side of the Eq. (\ref{dilaton
equation(I)}). It is a matter of calculation to show that (after
removing parts related to electromagnetic field)

\begin{equation}
T_{r}^{r}\propto \Lambda \left( \frac{b}{r_{+}}\right) ^{2\gamma }.
\end{equation}

Obtained relation indicates that although both dilaton field and rainbow
functions modified thermodynamical quantities, only the dilatonic part has
direct effect on the relation between cosmological constant and pressure.
Therefore, we use following analogy for studying the critical behavior of
the system
\begin{equation}
P=-\frac{\Lambda }{8\pi }\left( \frac{b}{r_{+}}\right) ^{2\gamma }.
\label{P}
\end{equation}

The conjugating quantity related to pressure is obtained through the use of
enthalpy
\begin{equation}
V=\left( \frac{\partial H}{\partial P}\right) _{T}.  \label{V}
\end{equation}

Since consideration of the cosmological constant extends our thermodynamical
phase space, the mass term plays the role of enthalpy. Therefore, by using
Eqs. (\ref{Mass}), (\ref{P}) and (\ref{V}), one can find modified volume of
these dilatonic black holes as
\begin{equation}
V=\frac{\mathcal{K}_{1,1}}{\mathcal{K}_{3,-1}g^{3}(\varepsilon
)f(\varepsilon )}r_{+}^{\frac{\mathcal{K}_{5,3}}{\mathcal{K}_{1,1}}%
}b^{2\gamma }.  \label{Volume}
\end{equation}

Clearly, the volume of these black holes is a function of both
rainbow functions and dilaton parameter. In other words, contrary
to some specific modified gravities, in this gravity, the volume
of the black hole is affected by the presence of rainbow and
dilaton gravities. Here, in order to have a positive and non-zero
volume, we find a restriction $-3<\alpha <3$. Since we are not
interested in negative values of the $\alpha$, we restrict
ourselves to $0<\alpha <3$.

It should be pointed out that due to the relation between volume of the
black hole and horizon radius, one is able to introduce specific volume for
these black holes which enables us to use horizon radius instead of volume
in following calculations. So, the pressure is given by
\begin{equation}
P=\frac{\mathcal{K}_{3,-1}g(\varepsilon )f(\varepsilon )\left( {\frac{b}{%
r_{c}}}\right) ^{2\gamma }r_{+}^{\frac{\mathcal{K}_{-1,1}}{\mathcal{K}_{1,1}}%
}b^{-2\gamma }}{2\mathcal{K}_{1,1}\mathcal{K}_{3,1}}\,{T}+\frac{\mathcal{K}%
_{3,-1}g^{2}(\varepsilon )\left[ f^{2}(\varepsilon )\mathcal{K}_{-1,1}{q}%
^{2}+{r}_{+}^{2}\right] }{8{r}_{+}^{4}{\pi }\mathcal{K}_{-1,1}\mathcal{K}%
_{3,1}}\left( {\frac{b}{r_{+}}}\right) ^{-2\gamma }.  \label{Pressure}
\end{equation}

In order to find a relation for calculating critical volume, hence critical
horizon radius, we use the concept of inflection point. In this method, one
uses
\[
\left( \frac{\partial P}{\partial r_{+}}\right) _{T}=\left( \frac{\partial
^{2}P}{\partial r_{+}^{2}}\right) _{T}=0,
\]%
to find critical horizon radius which in case of this thermodynamical system
is
\begin{equation}
r_{c}=qf(\varepsilon )\sqrt{\mathcal{K}_{3,1}\mathcal{K}_{2,1}},  \label{rc}
\end{equation}%
which will lead to following critical temperature and pressure
\begin{eqnarray}
T_{c} &=&\frac{\mathcal{K}_{1,1}g(\varepsilon )r_{c}^{\frac{\mathcal{K}%
_{1,-1}}{\mathcal{K}_{1,1}}}b^{2\gamma }}{f^{3}(\varepsilon ){q}^{2}{\pi }%
\mathcal{K}_{1,-1}\mathcal{K}_{2,1}^{2}\mathcal{K}_{3,1}^{2}}\left( \frac{b}{%
r_{c}}\right) ^{^{-4\gamma }},  \label{Tc} \\
&& \\
P_{c} &=&\,\frac{g^{2}(\varepsilon )\mathcal{K}_{3,-1}\left( {\frac{b}{r_{c}}%
}\right) ^{-2\gamma }}{8\pi f^{2}(\varepsilon ){q}^{2}\mathcal{K}_{3,1}%
\mathcal{K}_{2,1}}.
\end{eqnarray}

It is worthwhile to mention that the restriction that was observed
was originated only from dilatonic part of the solutions. In other
words, we have no restriction on values that charge and rainbow
functions can acquire and our system is only thermodynamically
restricted by dilaton parameter.

Using obtained critical values, one can find the following
critical ratio
\begin{equation}
\frac{P_{c}r_{c}}{T_{c}}=\frac{qg(\varepsilon )f^{2}(\varepsilon )\mathcal{K}%
_{-1,1}\mathcal{K}_{3,1}^{1/2}}{8\mathcal{K}_{1,1}\mathcal{K}_{2,1}^{1/2}}%
\left( {\frac{b}{r_{c}}}\right) ^{2\gamma }r_{c}^{\frac{\mathcal{K}_{1,-1}}{%
\mathcal{K}_{1,1}}}b^{2\gamma },  \label{ratio}
\end{equation}%
which shows that this critical ratio was modified due to the presence of
dilaton field as well as rainbow functions. It is worthwhile to mention that
critical horizon radius depends only on one of the rainbow functions whereas
the other critical values and also the ratio $\frac{P_{c}r_{c}}{T_{c}}$ are
functions of both of them. In addition, it is notable that in the absence of
dilaton field ($\alpha =0$) and low energy limit ($f(\varepsilon
)=g(\varepsilon )=1$), Eq. (\ref{ratio}) reduces to the usual universal
ratio in four dimensional Einstein gravity \cite{CosmP5,CosmP9}.

Next, using the renewed role of the total mass of the black holes, we have
Gibbs free energy as
\begin{equation}
G=H-TS=M-TS,  \label{G}
\end{equation}%
which by using Eqs. (\ref{temp}), (\ref{entropy}), (\ref{Mass}) and (\ref{P}%
) will be
\begin{equation}
G=\frac{\mathcal{K}_{1,1}^{2}\,{b}^{2\gamma }{r}_{+}^{{\frac{{\alpha }^{2}+3%
}{\mathcal{K}_{1,1}}}}}{2\mathcal{K}_{-3,1}f(\varepsilon )g^{3}(\varepsilon )%
}P+\frac{\mathcal{K}_{3,1}f(\varepsilon )\left( {\frac{b}{r_{c}}}\right)
^{-2\gamma }r_{c}^{-\frac{\mathcal{K}_{1,3}}{\mathcal{K}_{1,1}}}b^{2\gamma }%
}{16\pi g(\varepsilon )}{q}^{2}+\frac{\left( {\frac{b}{r_{c}}}\right)
^{-2\gamma }r_{c}^{\frac{\mathcal{K}_{1,-1}}{\mathcal{K}_{1,1}}}b^{2\gamma }%
}{16\pi f(\varepsilon )g(\varepsilon )}.
\end{equation}

\begin{figure}[tbp]
$%
\begin{array}{ccc}
\epsfxsize=5.5cm \epsffile{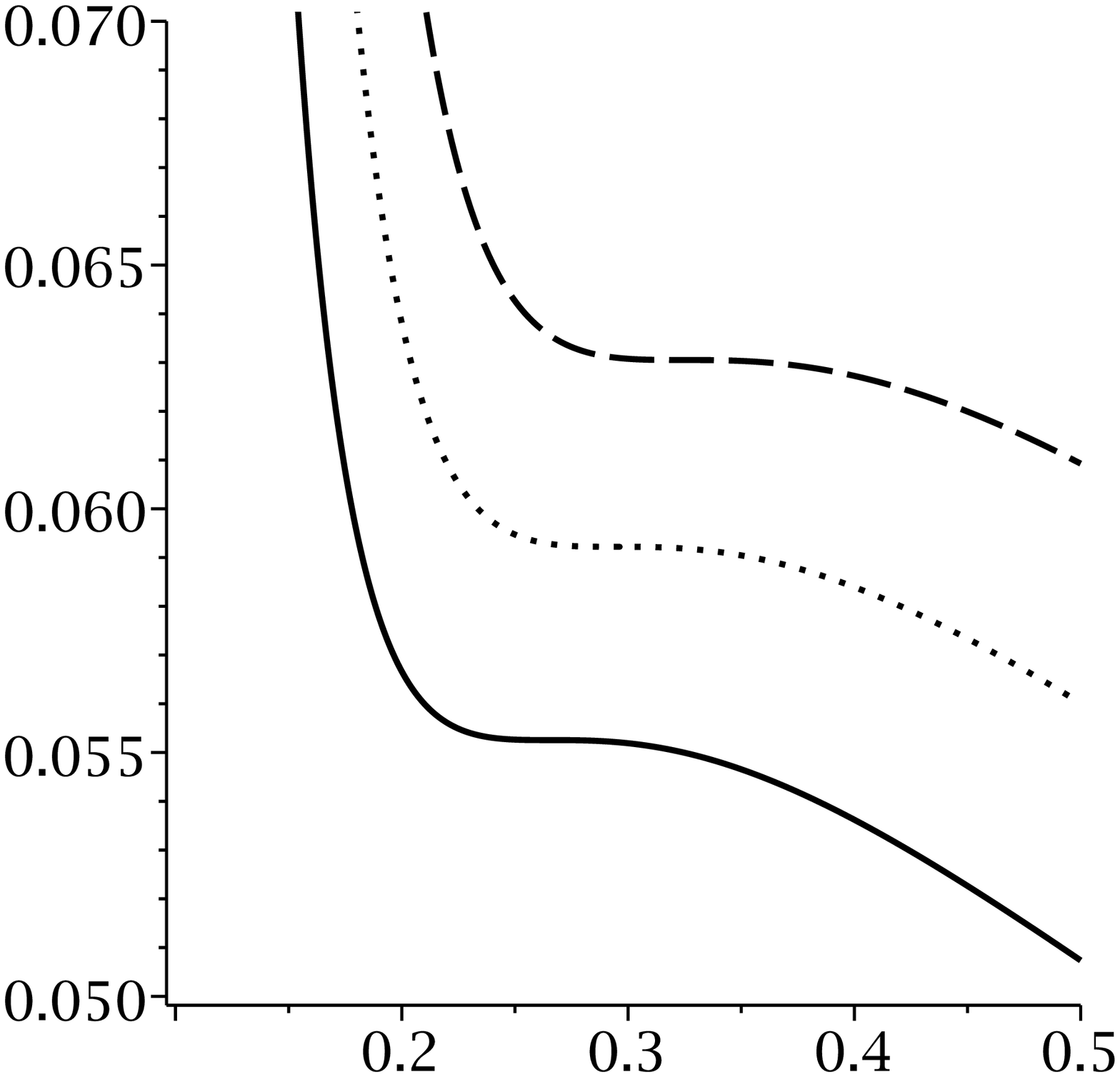} & \epsfxsize=5.5cm %
\epsffile{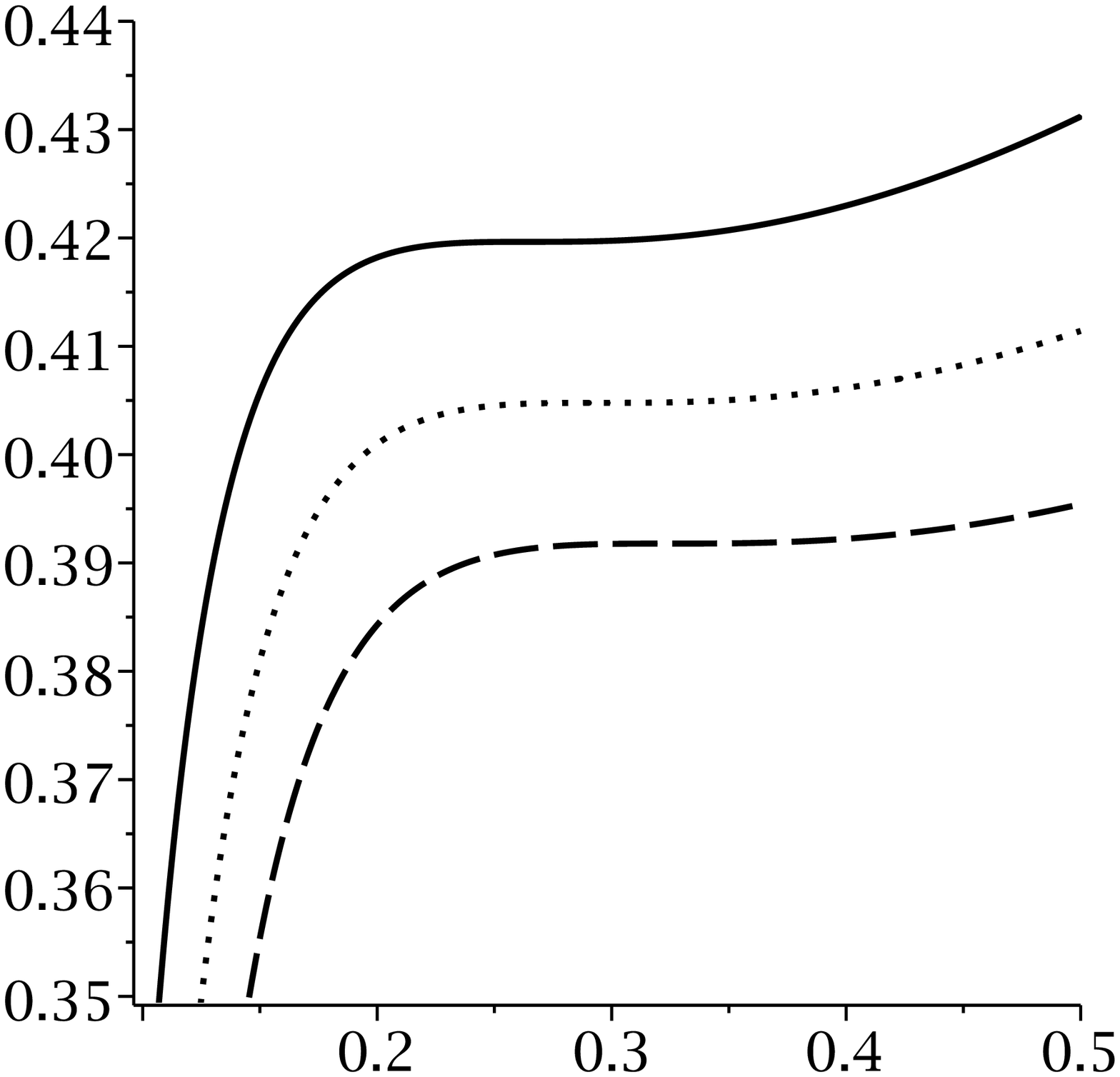} & \epsfxsize=5.5cm %
\epsffile{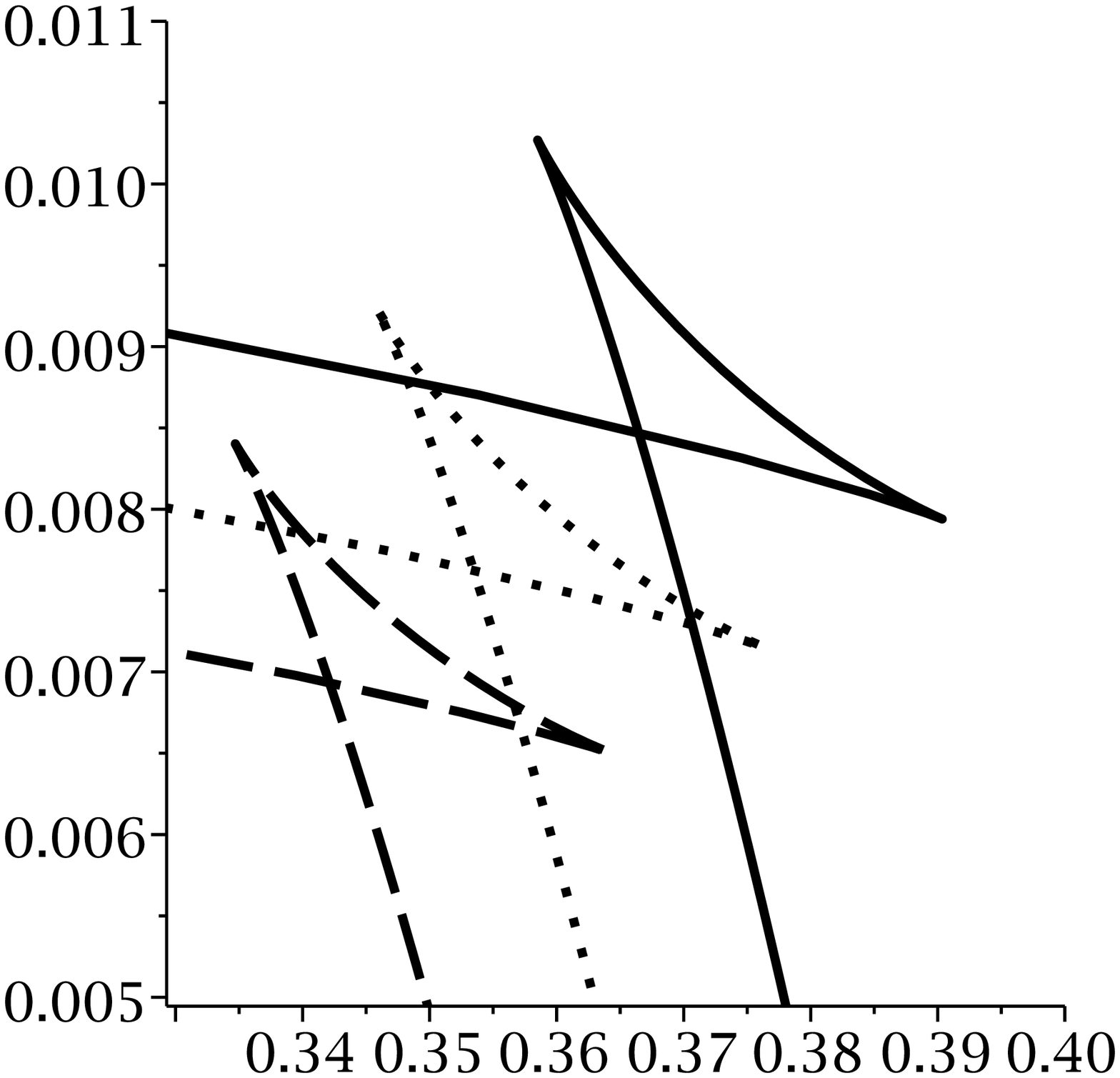}%
\end{array}
$%
\caption{ $P-r_{+}$ (left), $T-r_{+}$ (middle) and $G-T$ (right) diagrams
for $q=0.1$, $b=1$, $\protect\alpha=0.7$, $g(\protect\varepsilon)=f(\protect%
\varepsilon)=0.9$ (continuous line), $g(\protect\varepsilon)=f(\protect%
\varepsilon)=1$ (dotted line) and $g(\protect\varepsilon)=f(\protect%
\varepsilon)=1.1$ (dashed line). \newline
$P-r_{+}$ diagram for $T=T_{c}$, $T-r_{+}$ diagram for $P=P_{c}$ and $G-T$
diagram for $P=0.5P_{c}$. }
\label{Fig4}
\end{figure}

\begin{figure}[tbp]
$%
\begin{array}{ccc}
\epsfxsize=5.5cm \epsffile{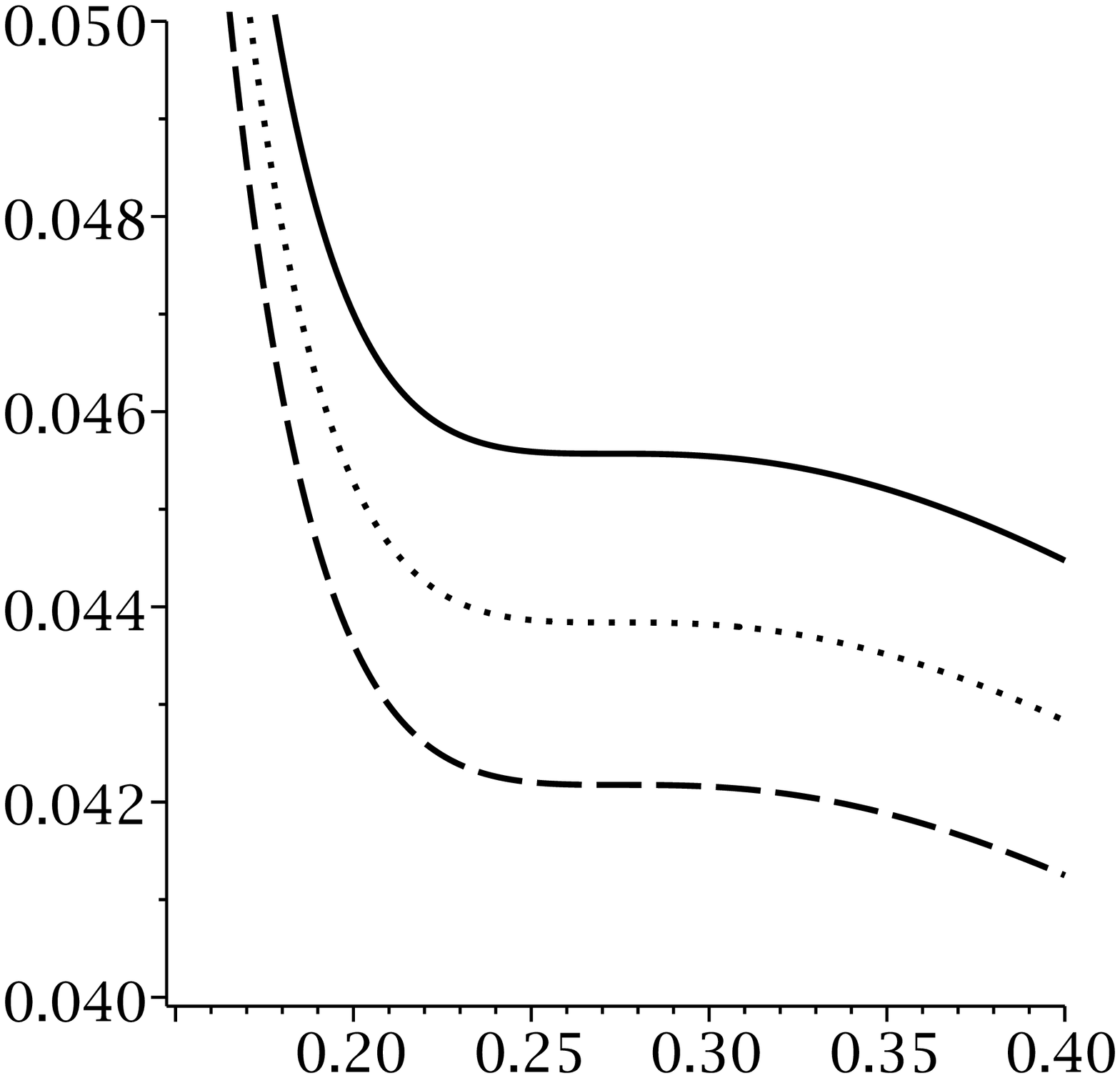} & \epsfxsize=5.5cm %
\epsffile{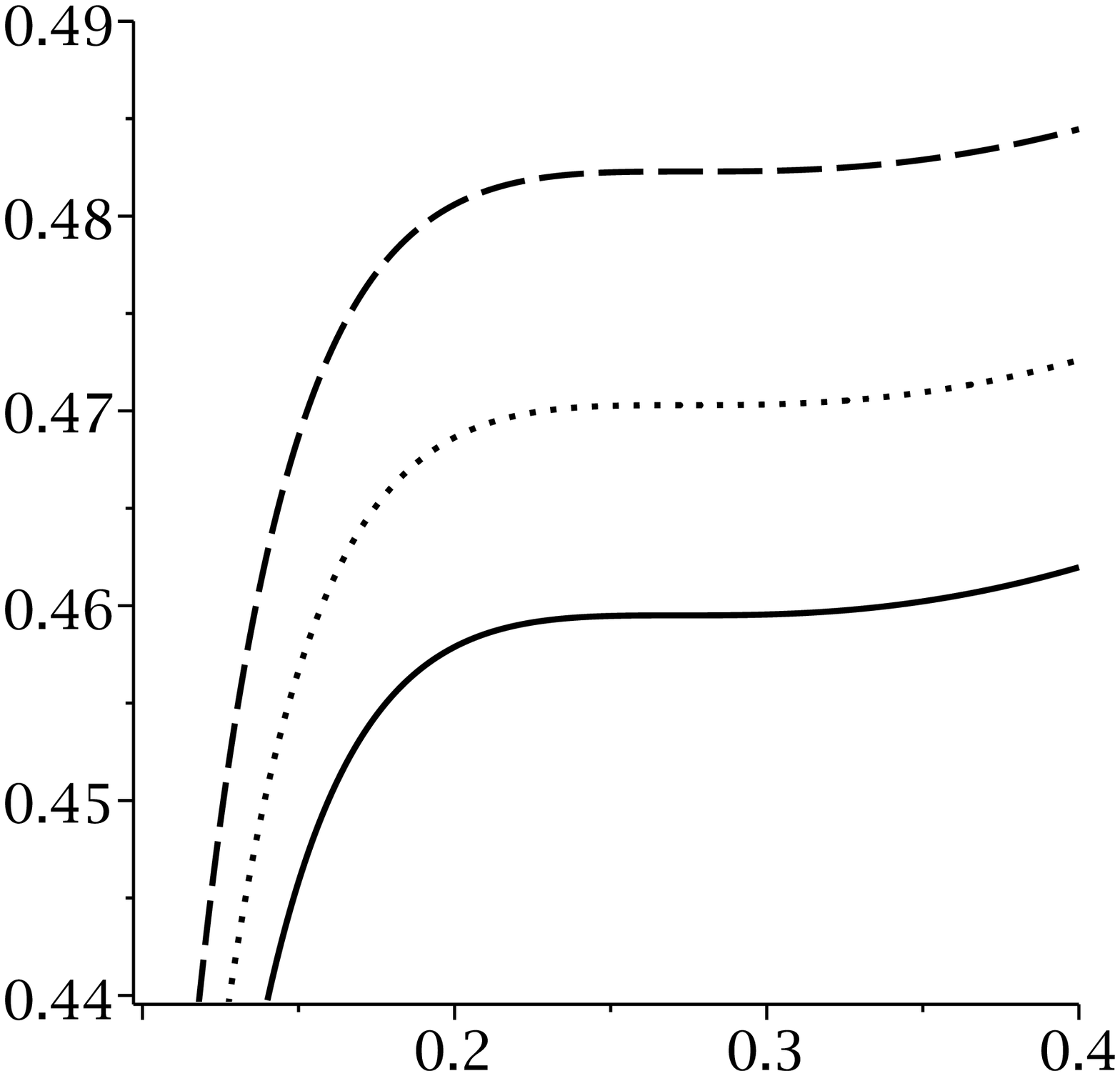} & \epsfxsize=5.5cm %
\epsffile{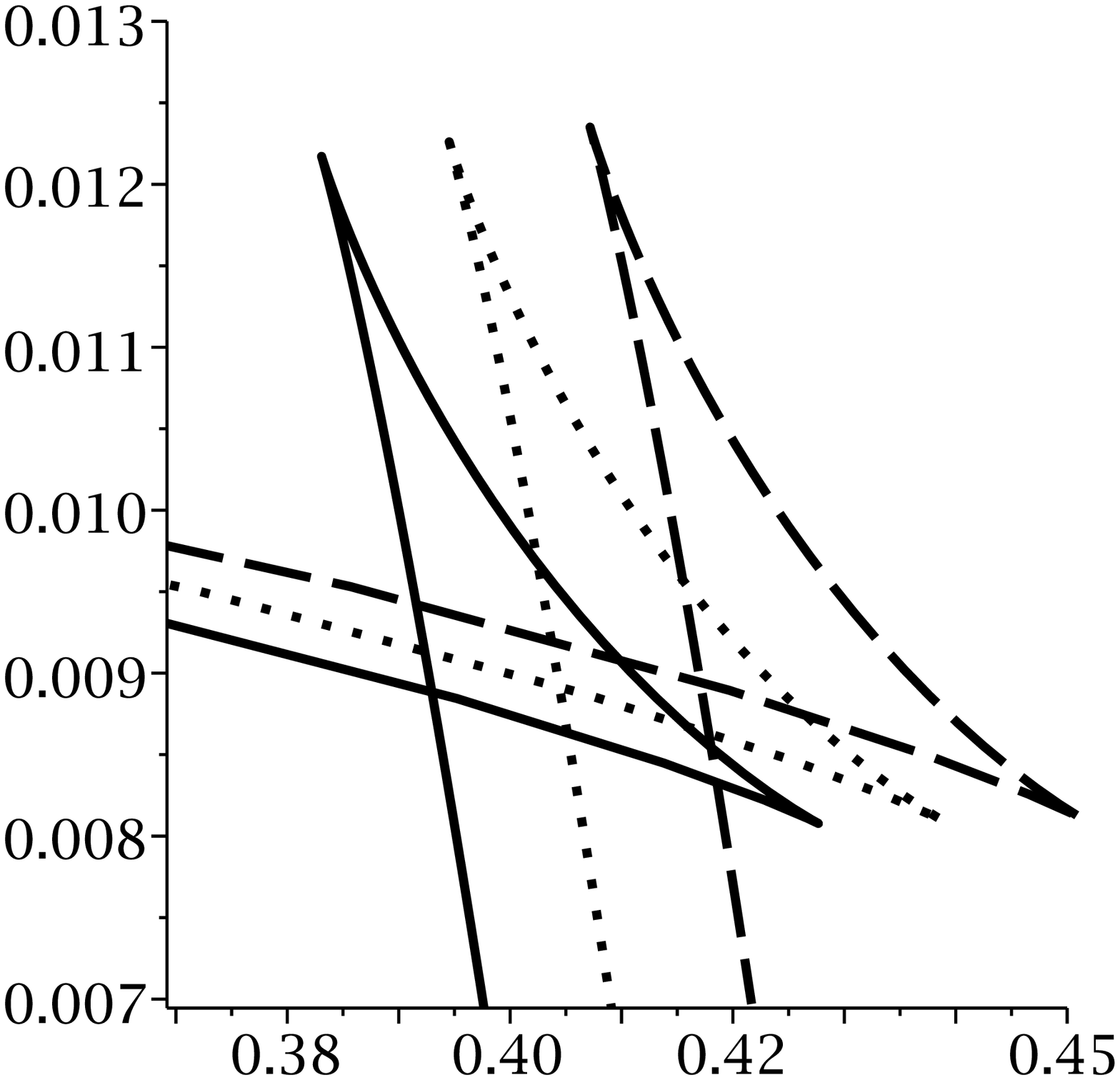}%
\end{array}
$%
\caption{ $P-r_{+}$ (left), $T-r_{+}$ (middle) and $G-T$ (right) diagrams
for $q=0.1$, $b=1$, $g(\protect\varepsilon)=f(\protect\varepsilon)=0.9$, $%
\protect\alpha=0.75$ (continuous line), $\protect\alpha=0.76$ (dotted line)
and $\protect\alpha=0.77$ (dashed line). \newline
$P-r_{+}$ diagram for $T=T_{c}$, $T-r_{+}$ diagram for $P=P_{c}$ and $G-T$
diagram for $P=0.4P_{c}$. }
\label{Fig5}
\end{figure}

\begin{figure}[tbp]
$%
\begin{array}{ccc}
\epsfxsize=5.5cm \epsffile{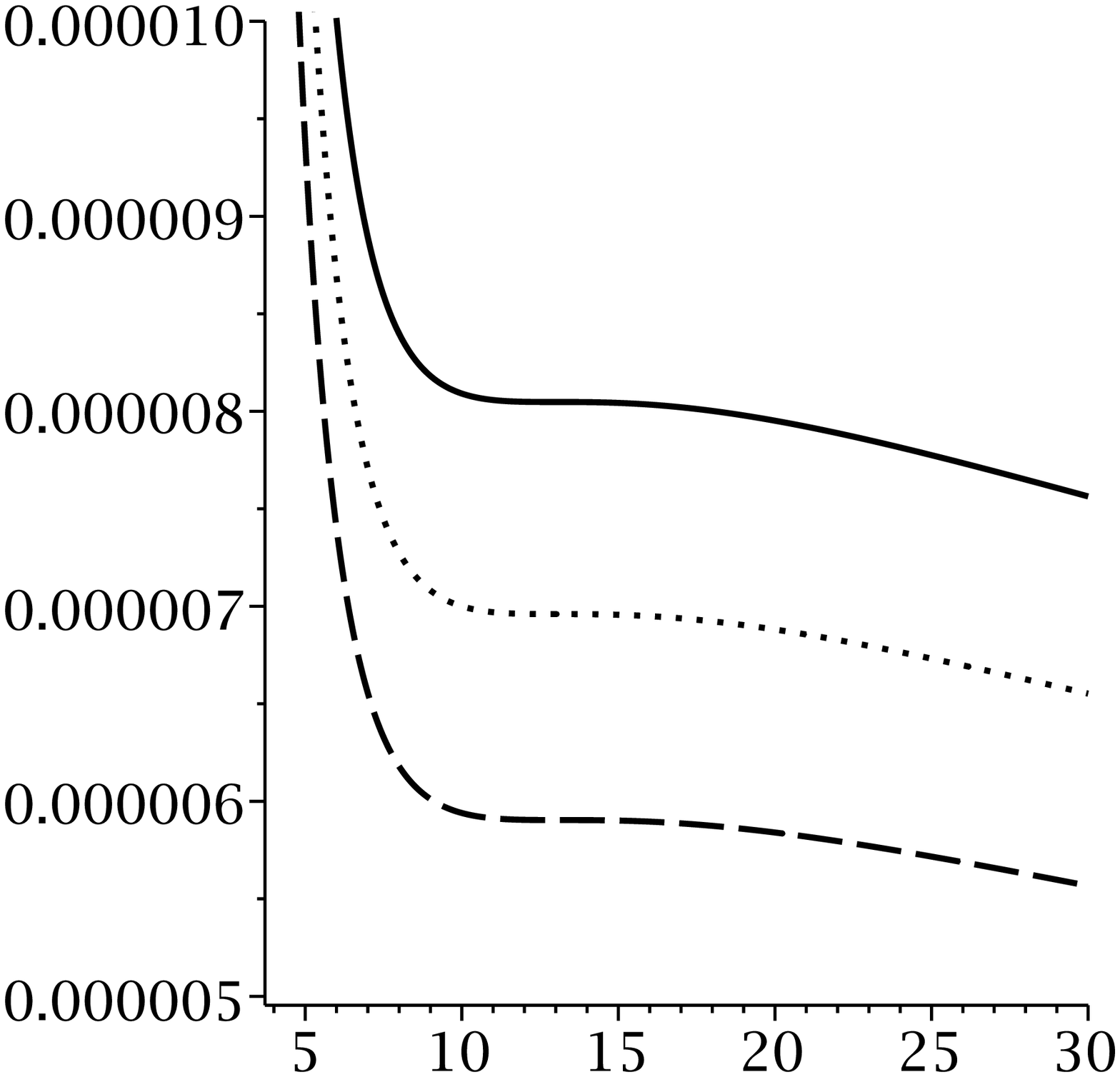} & \epsfxsize=5.5cm %
\epsffile{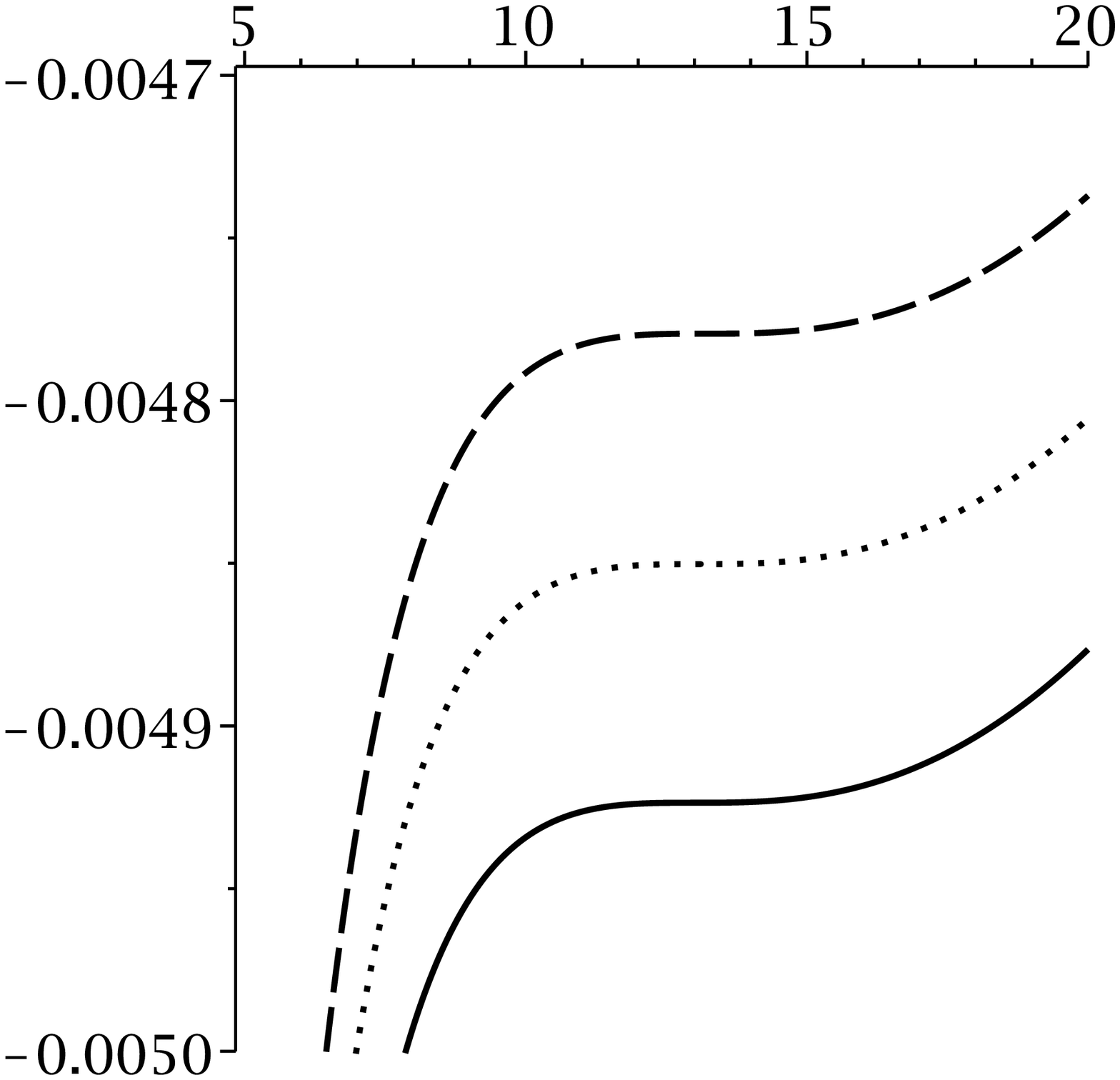} & \epsfxsize=5.5cm \epsffile{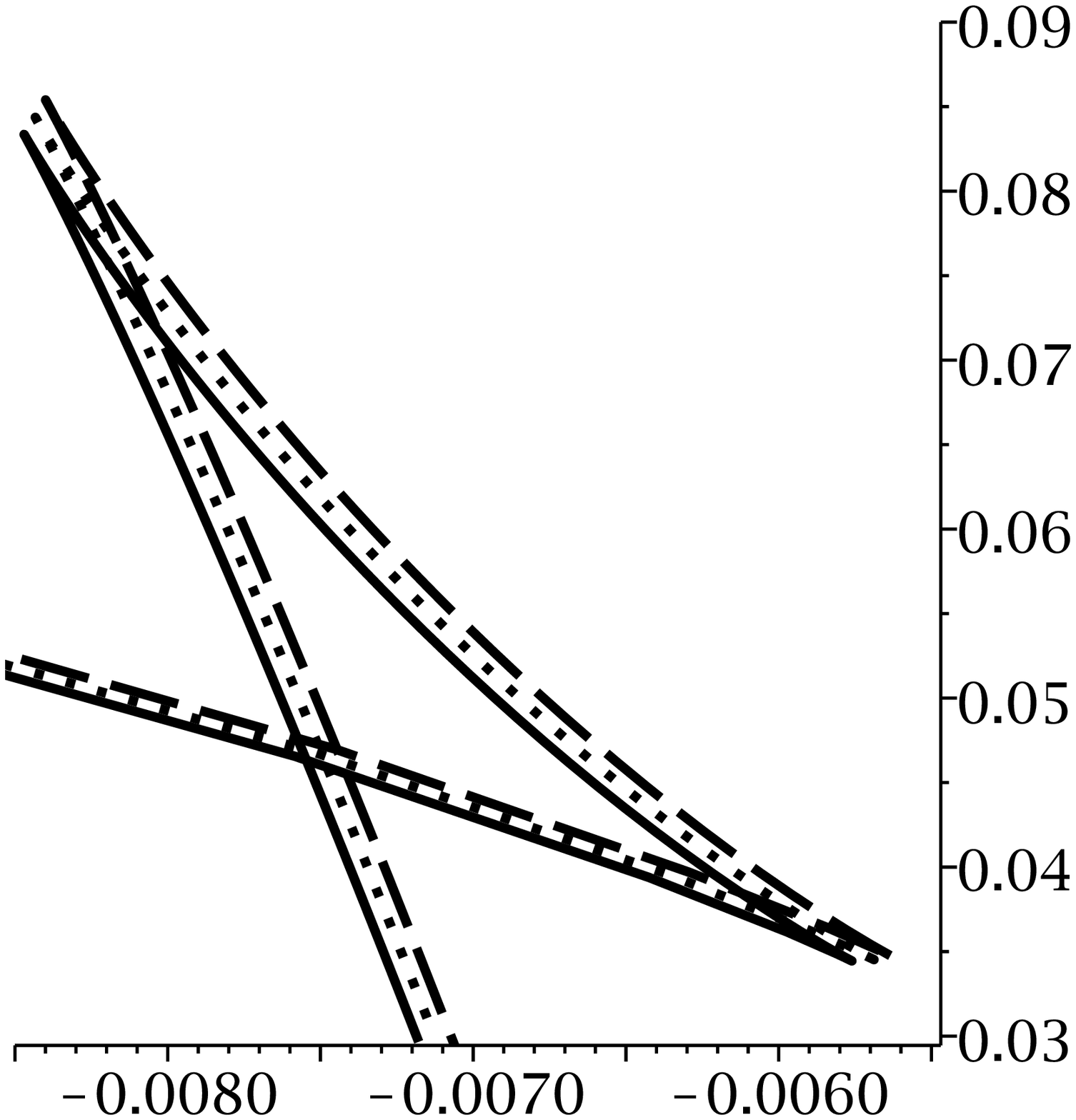}%
\end{array}
$%
\caption{ $P-r_{+}$ (left), $T-r_{+}$ (middle) and $G-T$ (right) diagrams
for $q=1$, $b=20$, $g(\protect\varepsilon)=f(\protect\varepsilon)=2.5$, $%
\protect\alpha=0.75$ (continuous line), $\protect\alpha=0.76$ (dotted line)
and $\protect\alpha=0.77$ (dashed line). \newline
$P-r_{+}$ diagram for $T=T_{c}$, $T-r_{+}$ diagram for $P=P_{c}$ and $G-T$
diagram for $P=0.5P_{c}$. }
\label{Fig6}
\end{figure}

In order to see whether obtained critical values represent a second order
phase transition, we study phase diagrams ($P-r_{+}$, $T-r_{+}$ and $G-T$
diagrams) in Figs. \ref{Fig4}-\ref{Fig6}.

It is evident that for specific values of different parameters, a second
order phase transition is observed for obtained critical values (see Figs. %
\ref{Fig4} and \ref{Fig5}). The critical pressure (left panels of Figs. \ref%
{Fig4} and \ref{Fig5}), temperature and subcritical isobars
(middle panels of Figs. \ref{Fig4} and \ref{Fig5}), energy of
different phases and size of swallow-tails (right panels of Figs.
\ref{Fig4} and \ref{Fig5}) are functions of gravity's rainbow and
dilaton parameter. The effects of rainbow functions and dilaton
parameter on critical values are different from each other
(compare Figs. \ref{Fig4} with \ref{Fig5}).

Interestingly, for a set of values, it is possible to obtain positive
critical pressure and horizon radius whereas the temperature is negative.
The plotted diagrams for these cases show a normal critical behavior in $%
P-r_{+}$ diagram (left panel of Fig. \ref{Fig6}) whereas in $G-T$ diagram an
abnormal behavior is observed (right panel of Fig. \ref{Fig6}). In case of $%
T-r_{+}$ diagram, also normal critical behavior is observed except that this
behavior is located in negative temperature.

\begin{figure}[tbp]
$%
\begin{array}{ccc}
\epsfxsize=5.5cm \epsffile{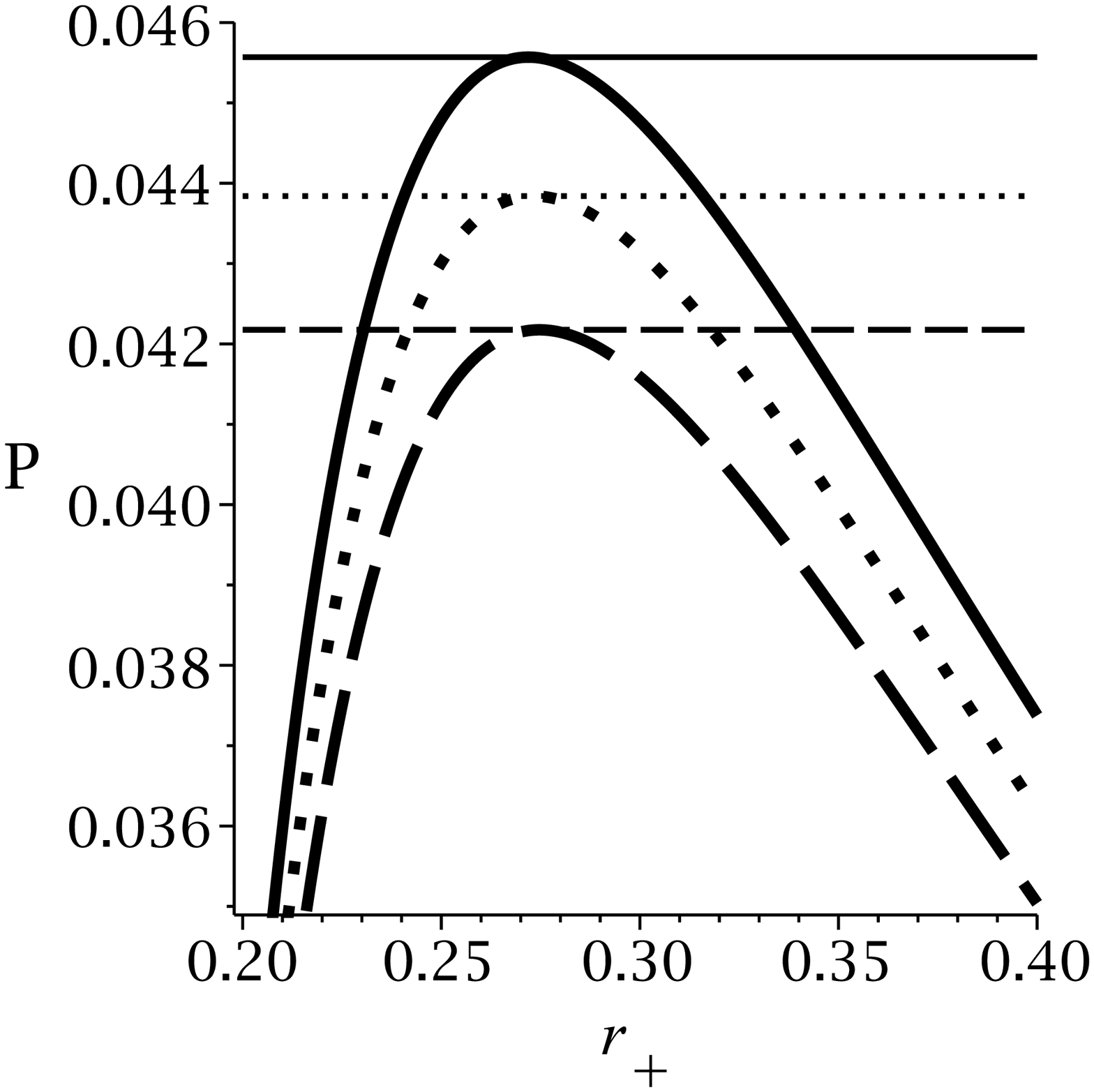} & \epsfxsize=5.5cm %
\epsffile{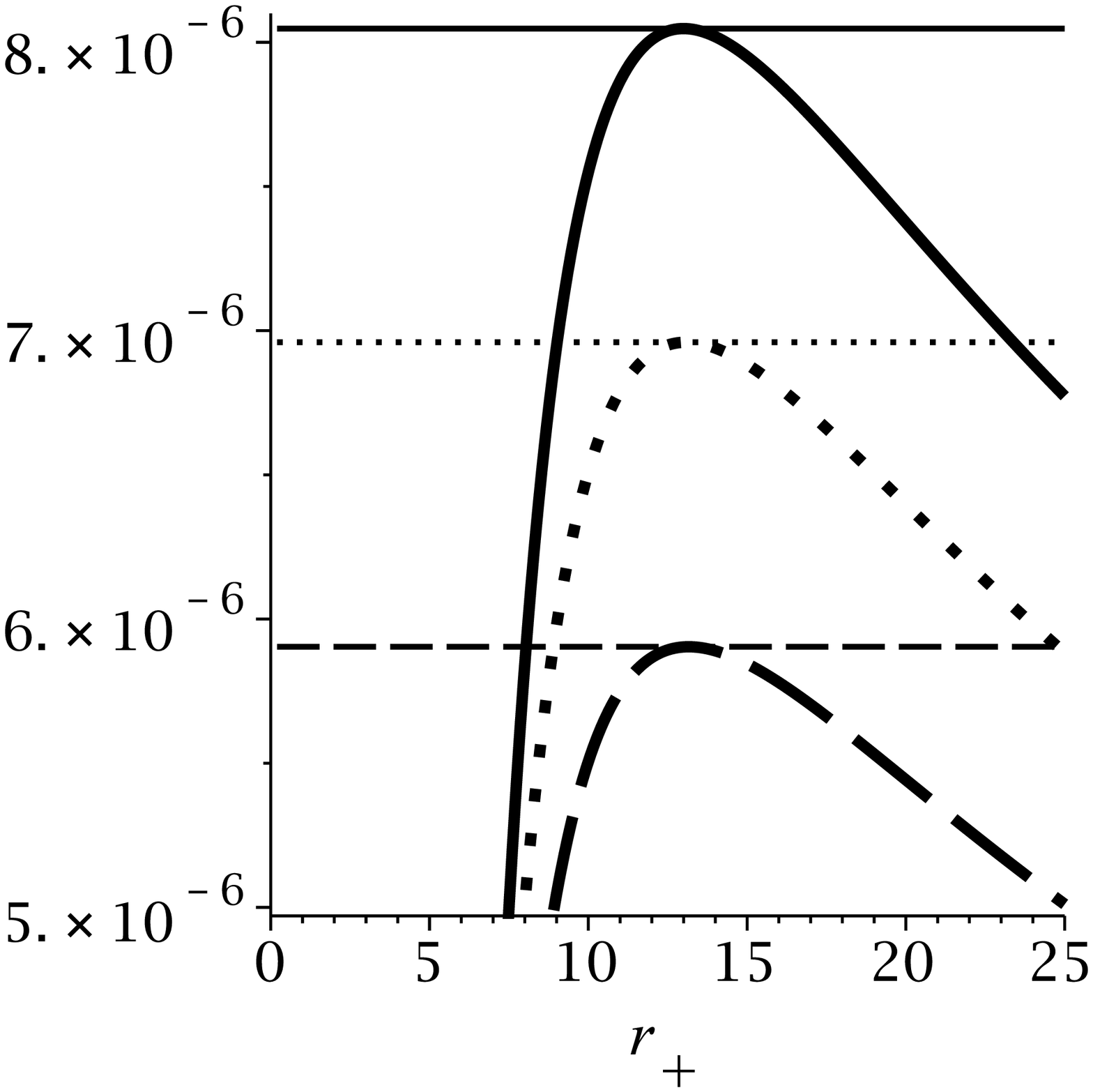} & \epsfxsize=5.5cm \epsffile{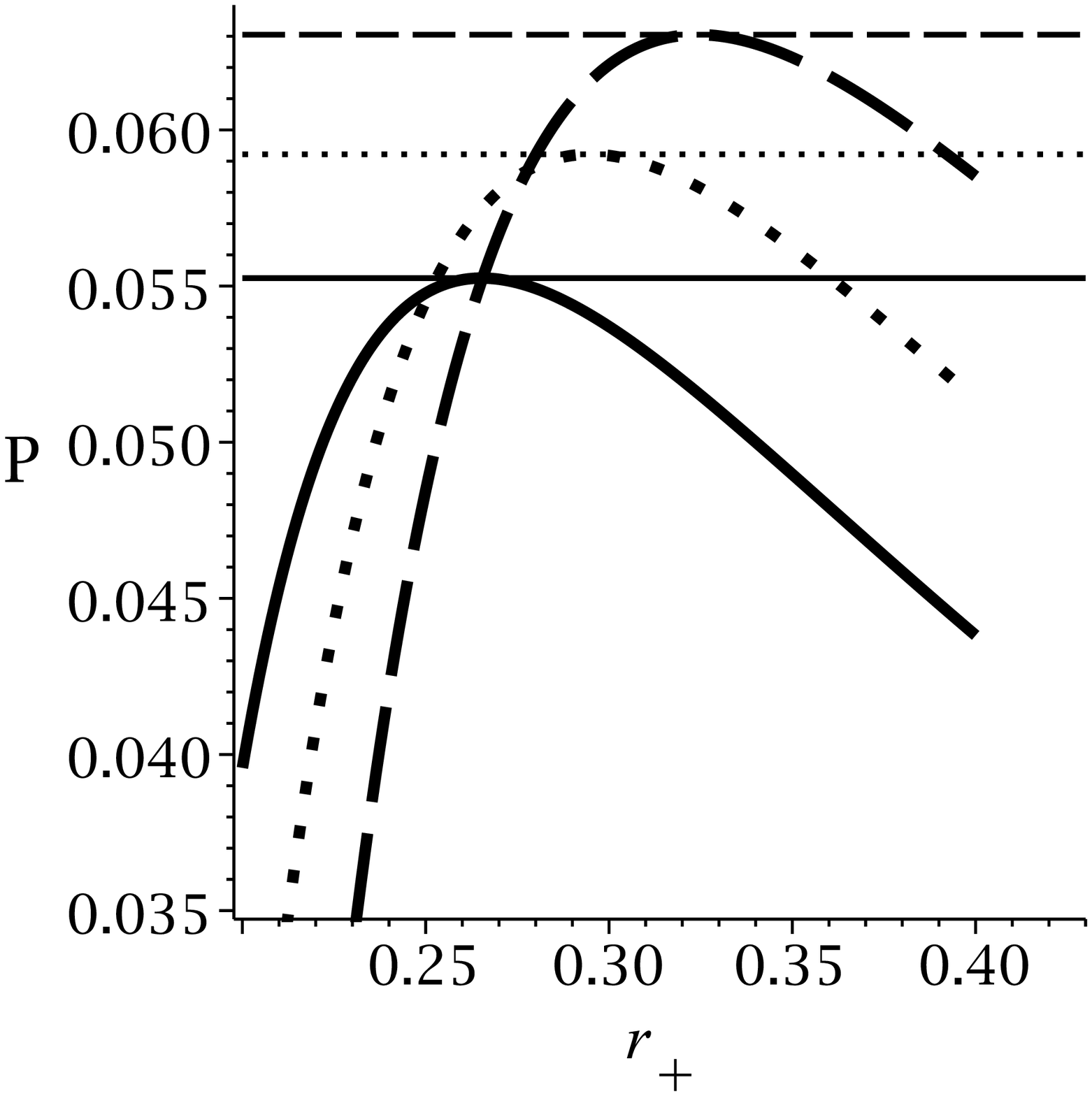}%
\end{array}
$%
\caption{$P$ versus $r_{+}$ diagrams for $q=0.1$, $b=1$ and $k=1$.
\newline
left panel: $g(\protect\varepsilon)=f(\protect\varepsilon)=0.9$ and $\protect%
\alpha=0.75$ (bold continues line), $P=0.0455693$ (continues line), $\protect%
\alpha=0.76$ (bold dotted line), $P=0.0438404$ (dotted line), $\protect\alpha%
=0.77$ (bold dashed line), $P=0.0421757$ (dashed line). \newline
middle panel: $g(\protect\varepsilon)=f(\protect\varepsilon)=0.9$ and $%
\protect\alpha=1.65$ (bold continues line), $P=0.0000080$ (continues line), $%
\protect\alpha=1.66$ (bold dotted line), $P=0.0000069$ (dotted line), $%
\protect\alpha=1.67$ (bold dashed line), $P=0.0000059$ (dashed
line).
\newline
right panel: $\protect\alpha=0.7$ and $g(\protect\varepsilon)=f(\protect%
\varepsilon)=0.9$ (bold continues line), $P=0.0552557$ (continues line), $g(%
\protect\varepsilon)=f(\protect\varepsilon)=1$ (bold dotted line), $%
P=0.0592206$ (dotted line), $g(\protect\varepsilon)=f(\protect\varepsilon%
)=1.1$ (bold dashed line), $P=0.0630518$ (dashed line).}
\label{Fig7}
\end{figure}

\section{Phase transition points through heat capacity}

In this section, we will obtain critical points through a method
which was developed in Ref. \cite{CosmP14}. In this method, the
denominator of the heat capacity is employed to obtain an explicit
relation for thermodynamical pressure. Obtained relation may yield
a maximum(s) for pressure which is(are) critical pressure(s) in
which a second order phase transition takes place. This critical
pressure is exactly the same as that of obtained through the use
of phase diagrams.

Using Eqs. (\ref{heat}) and (\ref{P}) and solving the denominator
with respect to thermodynamical pressure will lead to following
explicit relation for pressure
\begin{equation}
P=\frac{\mathcal{K}_{-3,1}\left[ q^{2}f^{2}(\varepsilon )\mathcal{K}%
_{3,1}-r_{+}^{2}\right] g^{2}(\varepsilon )}{8\pi r_{+}^{4}\mathcal{K}_{3,1}%
\mathcal{K}_{1,1}}\left( \frac{b}{r_{+}}\right) ^{-2\gamma }.  \label{PR}
\end{equation}

It is evident that this relation is different from previously
obtained relation for pressure (Eq. \ref{Pressure}). Now, by using
values that are employed for plotting phase diagrams (Figs.
\ref{Fig4}-\ref{Fig6}), we plot following diagrams (Fig.
\ref{Fig7}). A simple comparison shows that the maximums of the
plotted diagrams are exactly where corresponding critical pressure
and horizon radius are located in phase diagrams. This shows that
these two approaches yield consisting picture regarding the
critical behavior of these black holes. On the other hand, plotted
diagram which corresponds to one with abnormal behavior (middle
panel of Fig. \ref{Fig7}) also represents the characteristic
behavior of the phase transition point. Therefore, in case of
these black holes, a phase transition occurs in the mentioned
critical point.

\section{Conclusion}

In this paper, we studied $4$-dimensional charged dilatonic black
holes in gravity's rainbow and their thermal stability conditions.
We obtained thermodynamical quantities such as temperature,
electric charge, entropy and total mass of the black holes. These
quantities were modified in gravity's rainbow and became energy
dependent.

Next, we conducted a study regarding physical/nonphysical black
holes (positivity/negativity of temperature) and thermal stability
of the solutions. It was pointed out that dominant factor in
studying these properties is the dilaton parameter. In other
words, these properties were highly sensitive to variation of
$\alpha$. Due to different factors of dilaton parameter, different
types of behavior were observed for temperature which put
restrictions on the solutions being physical. Observed behaviors
for the temperature were: a) two roots with maximum, b) an
increasing (a decreasing) function of horizon radius with one root
c) a negative definite function with one maximum.

The analyzed behaviors were, increasing function of horizon radius
with one root, increasing and decreasing function of horizon
radius with two roots, decreasing function of the $r_{+}$ with one
root located and being negative with one maximum located at
negative temperature.

As for the stability and phase transition, we found depending on the
behavior of the temperature, heat capacity could have phase transition and
stable state for larger values of horizon radius. In case of two roots for
temperature, interestingly, a phase transition of larger/smaller black hole
was observed. Finally, as for temperature being decreasing function of $%
r_{+} $, for physical solutions (positive temperature), unstable solutions
were observed. In other words, in this case, physical solutions are unstable.

Next, geometrical approach was employed to study the bound and
phase transition points of these black holes. It was demonstrated
that the Ricci scalars of the phase spaces of Weinhold, Ruppeiner
and Quevedo metrics, have divergencies which do not match with
mentioned points while the singular points of curvature scalar of
the HPEM coincide with roots and divergence points of the heat
capacity.

We also, studied the critical behavior of these black holes in
extended phase space. It was shown that the usual relation between
cosmological constant and thermodynamical pressure was modified
due to existence of dilaton gravity whereas such modification was
not seen for gravity's rainbow. On the contrary, it was shown that
volume of the black holes depends on both of these modifications.

Then, we showed that the critical values are related to the
rainbow functions as well as the dilaton parameter. In order to
have positive critical temperature, we found restrictions which
were purely dilatonic dependant. Therefore, one is not free to
choose any value for dilaton parameter.

In studying phase diagrams, two different behaviors were observed
for different diagrams, especially in $G-T$ diagrams.
Interestingly, although we observed an anomaly in these diagrams,
other corresponding phase diagrams presented usual thermodynamical
behavior around critical points. In other words, the observed
abnormal behaviors in phase diagrams present the existence of a
second order phase transition for these black holes.

Next, we used a new method which was introduced in Ref.
\cite{CosmP14}, for studying the critical behavior of the system.
This method is based on obtaining an explicit form for
thermodynamical pressure from denominator of the heat capacity. It
was seen that the maximum maximum of this relation is located at
the critical pressure and horizon radius in which second order
phase transition takes place. It was shown that in this method,
for irregular behavior which was observed in phase diagrams, also
a second order phase transition occurs. This indicates that these
points are phase transition point despite their abnormal behavior.

Finally, it is worthwhile to think about the physical
interpretation of abnormal behavior which was seen in this paper.
It is notable that, one can generalize obtained linear solutions
in this paper to nonlinear case of electrodynamics and investigate
the effects of nonlinearity \cite{Nonlinear}. In addition, one may
investigate the extended phase space and thermodynamic criticality
in higher order Lovelock-Maxwell gravity's rainbow as well as
Lovelock-nonlinear electrodynamics
\cite{HendiDehghani,HendiArshad}. These subjects are under
examination.

\begin{acknowledgements}
We thank the Shiraz University Research Council. This work has
been supported financially by the Research Institute for Astronomy
and Astrophysics of Maragha, Iran.
\end{acknowledgements}

\end{document}